\documentclass[11pt]{article}
\usepackage{amsmath,amsfonts,amssymb,amsthm,bbm,mathptm,cases}
\usepackage{graphicx,graphics,epsfig,subfigure}
\usepackage{graphicx}
\usepackage{booktabs}
\usepackage{threeparttable}
\usepackage{mathrsfs}
\usepackage{multirow}
\usepackage{appendix}
\usepackage{xcolor,color,soul,array,cite,float,version,times}
\newtheorem{remark}{Remark}
\numberwithin{equation}{section}
\numberwithin{figure}{section}
\numberwithin{table}{section}
\numberwithin{footnote}{section}
\theoremstyle{definition}

\theoremstyle{plain}

\theoremstyle{remark}

\newcommand{\ben}{\begin{eqnarray}}
\newcommand{\een}{\end{eqnarray}}
\newcommand{\bea}{\begin{array}}
\newcommand{\eea}{\end{array}}
\newcommand{\bes}{\begin{subequations}}
\newcommand{\ees}{\end{subequations}}
\newcommand{\bef}{\begin{figure}[H]}
\newcommand{\eef}{\end{figure}}
\newcommand{\bet}{\begin{tikzpicture}}
\newcommand{\eet}{\end{tikzpicture}}
\newcommand{\beq}{\begin{equation}}
\newcommand{\eeq}{\end{equation}}
\def\bena#1\eena{\begin{eqnarray}\begin{array}{l}#1\end{array}\end{eqnarray}}
\def\besl#1\eesl{\begin{subequations}\begin{align}#1\end{align}\end{subequations}}

\newcommand{\parl}[2]{\ensuremath{\frac{\partial #1}{\partial #2}}}

\def\inc(#1){\includegraphics[height=3 cm]{pics/#1}}

\def\bR{\mathbf{R}}
\def\bU{\mathbf{U}}

\def\bv{\mathbf{v}}

\def\bn{\mathbf{n}}
\def\bs{\mathbf{s}}
\def\b0{\mathbf{0}}

\def\bx{\mathbf{x}}

\def\bC{\mathbf{C}}

\def\bA{\mathbf{A}}
\def\bI{\mathbf{I}}

\def\bM{\mathbf{M}}
\def\bQ{\mathbf{Q}}

\def\bP{\mathbf{P}}

\begin{document}

\title{Toward a Physical Interpretation of Phase Field Models with Dynamic Boundary Conditions }
\author{
{Xiaobo Jing}\footnote{xiaobo@seu.edu.cn, School of Mathematics, Southeast University, Nanjing 210096, Jiangsu Province, P.R. China.},
{Qi Wang}\footnote{qwang@math.sc.edu, Department of Mathematics,
 University of South Carolina, Columbia, SC 29028, USA.}
}
\date{\today}
\maketitle
\begin{abstract}
In recent decades, considerable research has been devoted to partial differential equations (PDEs) with dynamic boundary conditions. However, the physical interpretation of the parameters involved often remains unclear, which in turn limits both theoretical analysis and numerical computation. For instance, the Robin boundary condition used in thermodynamically consistent models with dynamic boundary conditions has been misinterpreted as representing a chemical reaction, or has been generalized in an unjustified manner in numerous works. In this paper, we treat the bulk and surface as a closed system and develop  thermodynamically consistent phase field models to clarify the physical meaning of parameters in governing equations and boundary conditions, with particular focus on material and energy exchange between the bulk and surface by connecting it with the nanothermodynamics. Firstly, we commence with the mass and volume conservation law in the close system and  elucidate the physical interpretation of the Robin boundary condition, demonstrating that the relevant parameters are connected to the system's characteristic length scale and play a crucial role on the exchanging of material and energy. Furthermore, our analysis justifies the physical necessity for the phase variable in the bulk to differ from that on the surface. Secondly, we construct four more general models capable of describing both irreversible and irreversible-reversible coupling processes using the generalized Onsager principle. Thirdly, we reveal that both conservation and dissipation laws simultaneously determine the mobility operator and free energy, which are two dual variables.  Finally, we perform structure-preserving numerical simulations to systematically investigate how reversible processes and characteristic length affect pattern formation.
\end{abstract}

\section{Introduction}

\noindent \indent Building on our prior works \cite{jing2022thermodynamically, jing2023CMS}, which established a general framework for thermodynamically consistent phase field models with dynamic boundary conditions, we showed that all existing models of this type, including those with static boundary conditions, are special cases of our framework. However, key conceptual ambiguities remain in the field. A major source of confusion lies in the physical interpretation of the Robin boundary condition, expressed as $\alpha f_m = \beta \mu_s - \mu_b$, which describes the flux $f_m$ between the bulk and surface is governed by the bulk chemical potentials $\mu_b$, surface chemical potential $\mu_s$, the relaxation parameter $\alpha$ and another parameter $\beta$. To date, most interpretations—such as viewing it as a chemical reaction in \cite{knopf2021phase, wu2022review}—rely on mathematical rather than physical reasoning. Crucially, $\mu_b(\bs)$ here represents the limitation of the bulk chemical potential $\mu_b(\bx)$ to the surface coordinates $\bs$, not an physical variable. Therefore, interpreting it as a reaction is physically unsound and the Robin boundary condition should not be overgeneralized as in our previous work \cite{jing2022thermodynamically}. In this paper, we rederive the models from fundamental physical laws to clarify the true meaning of such boundary conditions.

When discussing a system in the material universe, we need to discuss the exchange of material and energy between the system and its environment/surrounding. In this case, the surface of system is so important that determines which statistical ensemble and thermodynamic characteristic function should be chosen. If the surface is peeled into the environment, the mass conservation law does not hold anymore for such an open system, the grand Canonical ensemble  and grand potential should be used. In this paper, we consider a system with bulk and surface as a closed system, which satisfies the conservation laws of mass. Unlike macroscopic systems, the thermodynamic properties of small systems with finite size depend on the specific experimental "environmental" independent variable. This variable is typically formulated as the subdivision potential in theoretical frameworks \cite{hill1962thermodynamics,hill1964thermodynamics}. In such systems with finite size, the surface of the system can significantly affects the bulk of the system and vice versa. For example, surface thermodynamics in nanomaterials depends on the size of nanoparticles\cite{li2016size}, the biomoleculer in the cytoplasm can improve the phase separation on the membrane \cite{zhao2021thermodynamics}.  Hong Qian discussed the the thermodynamics for such a small rapidly stirred biochemical reaction system \cite{qian2012hill}, based on the foundamental work of Terry Hill\cite{hill1962thermodynamics,hill1964thermodynamics}. Wei Dong introduced the surface tension on the nano system and discussed thermodynamics of a small system in \cite{dong2021thermodynamics}.

The relationship of thermodynamics and partial differential equation is not mutually independent. In applied mathematics field, the models satifying the laws of thermodynamics and other physical laws are named as the thermodynamically consistent model. Based on these models, there are many works on PDE analysis and numerical algorithm \cite{jing2020linear,liu2019energetic,knopf2019convergence,knopf2020phase,zhao2018general}. The bridge of thermodynamics and thermodynamically consistent PDE can be build by generalized Onsager principle \cite{wang2020generalized}, Onsager variational principle \cite{doi2011onsager}, energy variational approach \cite{hyon2012energy} and other modeling theories. The key idea of these model theories origins from Onsager's work \cite{onsager1931reciprocal1,onsager1931reciprocal2}. In this paper, we used the generalized Onsager principle, which is more creative and simplier than other theories. This methodology is simple, the principle assumes the generalized flux and the generalized force (or say chemical potential) are linked by the mobility operator.  Such relation will guarantee the dynamics obeys the second law of thermodynamics. Based on this, a lot of thermodynamically consisitent PDE model are developed for hydrodynamical system, multi-phase particle flow system, copolymer melts system and so on\cite{sun2025cross,hong2025hybrid,shen2022thermodynamically}.
 In many works on dynamic boundary conditions, the exchanges of material and energy between the system and its environment are ignored by setting period or static boundary condition, thus the surface dynamics is also ignored. In other words, most of these works focused on the dynamics in the bulk, which is coincide with the assumption of thermodynamic limit of classical thermodynamics by setting bulk as the whole system. However, such assumption will not hold when the shape, dynamics and energy of boundary/surface can affect the bulk. Bob Eisenberg also regarded that the statistical mechanics withoutspatial bounds is impossible as well as imperfect in \cite{Eisenberg2022}. So if the interaction of boundary and bulk is significant, then the system should involve the bulk and surface/boundary.  In this paper, we will focus on a special case, in which there are exchange of both material and energy between the bulk and surface in the system and there are no exchange of material between the system and its environment. We also assume the surface of the system is fixed and the temperature is constant, thus we can use the Canonical essemble to describe the system. In this paper, we neglect the convection process and osmosis to simplify the transport processes. 

In this paper, we have elucidated the physical meaning of the boundary conditions and all relevant parameters, clarified the relationship and distinction between the order parameters in the bulk and on the surface, and developed more reasonable and general thermodynamically consistent phase field models. We have also demonstrated the advantage of the generalized Onsager principle in describing irreversible-reversible processes. Together with our previous works \cite{jing2022thermodynamically, jing2023CMS}, we show that all previously published phase-field models with dynamic boundary conditions are special cases of the new models proposed here. Furthermore, through asymptotic analysis, most models with static boundary conditions can also be viewed as special cases of those with dynamic boundary conditions. Since the corresponding analysis has been thoroughly presented in \cite{jing2022thermodynamically, jing2023CMS}, it will not be repeated here. Finally, building on a solid statistical physics understanding of boundary conditions, the PDE system with dynamic boundary conditions can be naturally extended to hydrodynamic settings.

This paper is structured as follows. In Section 2, we analyze the exchange of mass and energy between the bulk and surface by introducing Robin boundary conditions, based on the conservation law and the second law of thermodynamics. In Section 3, we derive the governing equations and the associated energy dissipation using the generalized Onsager principle. In Section 4, we develop four thermodynamically consistent phase-field models via this principle, demonstrating that the bulk-surface exchange critically influences the mobility operator. Section 5 presents numerical simulations that highlight the differences between purely irreversible and mixed irreversible-reversible transport systems, revealing how the characteristic length or say parameter $\beta$ regulates mass and energy exchange. Finally, conclusions are summarized in Section 6.

\section{The exchange of material, conservation law and dissipation law}
\noindent \indent This section presents the derivation of two phase field models with a volume fraction and a area fraction as the phase variables to describe material exchange between the bulk and the surface. In Section 2.1, the first model is formulated using the volume fraction and area fraction as the order parameters, enforcing total mass conservation. Section 2.2 develops the second model, which incorporates total volume conservation with the same order parameters. Finally, the energy dissipation law for the models is considered in Section 2.3.
\subsection{Mass conservation law with the exchange of material between bulk and surface}
\noindent \indent  We firstly consider a binary fluid system, where material exchanges between bulk and surface.
In the bulk, the molecular number densities for the two component fluids are denoted by $n_b^{(i)}, i=1,2$ and the volume of one molecule are denoted by $v_i, i=1,2$. The corresponding mass is defined as $m_i$. So the mass density of $i^{th}$ species and the corresponding specific mass density (constant mass density of pure $i^{th}$ species) in the bulk is given by 
\bena
\rho_b^{(i)}=m_in_b^{(i)}, \quad \hat \rho_b^{(i)}=\frac{m_i}{v_i} .
\eena
Similarly, we can also introduce the molecular number densities on the surface and the effective area by $n_s^{(i)}$ and $s_i$, then the mass density and the corresponding specific mass density on the surface are given by 
\bena
\rho_s^{(i)}=m_in_s^{(i)}, \quad \hat \rho_s^{(i)}=\frac{m_i}{s_i} .
\eena 
We then define $\rho_b, \rho_s$ as the mass density of the fluid mixture in bulk and on the surface, $M_b$ and $M_s$ as the total mass in bulk and on the surface, where
\bena
M_b=\int_\Omega \rho_b d\bx=M_b^{(1)}+M_b^{(2)}, \quad
M_b^{(1)}=\int_\Omega \rho_b^{(1)} d\bx, \quad
M_b^{(2)}=\int_\Omega \rho_b^{(2)}  d\bx,\quad \rho_b=\rho_b^{(1)}+\rho_b^{(2)},\\
M_s=\int_{\partial \Omega} \rho_s d\bs=M_s^{(1)}+M_s^{(2)}, \quad M_s^{(1)}=\int_{\partial \Omega} \rho_s^{(1)} d\bs,
\quad M_s^{(2)}=\int_{\partial \Omega} \rho_s^{(2)} d\bs,\quad \rho_s=\rho_s^{(1)}+\rho_s^{(2)},
\eena
$\Omega$ is the bulk domain and $\partial \Omega$ is the surface. Now let us discuss the movement of fluid mixture in a fixed domain and neglect the convection effects. We assume the total mass for the same species of fluid in the system is constant, so we have
\bena
\frac{d(M_b^{(1)}+M_s^{(1)})}{dt}=\frac{d}{dt}\int_\Omega \rho_b^{(1)} d\bx+\frac{d}{dt}\int_{\partial \Omega} \rho_s^{(1)} d\bs
=\int_\Omega [\parl{\rho_b^{(1)}}{t} +\delta_{\partial \Omega} \parl{\rho_s^{(1)}}{t}]d\bx=0,\label{densityevo1}
\eena
where $\int_\Omega \delta_{\partial \Omega} f(\bx) d\bx=\int_{\partial \Omega} f(\bs) d\bs$ for any function $f(\bs)$.

Now let us introduce the order parameters $\phi_i$, $\psi_i$ to describe the volume fraction in the bulk and area fraction on the surface, respectively,
\bena
\phi_i=\frac{\rho_b^{(i)}}{\hat \rho_b^{(i)}}=n_b^{(i)}v_i,\quad 
\psi_i=\frac{\rho_s^{(i)}}{\hat \rho_s^{(i)}}=n_s^{(i)}s_i, \quad
\frac{\phi_i}{\psi_i}=\frac{n_b^{(i)}v_i}{n_s^{(i)}s_i},
\eena
which satisfy $\phi_1+\phi_2=1$, $\psi_1+\psi_2=1$.
\begin{remark}
Based on the definitions of order parameters, we can find $\psi_i$ are not necessarily the same as $\phi_i$ confined to the surface.	
\end{remark}  Substituting above two order parameters into \eqref{densityevo1}, we can get the transport equation for $\phi_1$ and $\psi_1$ is given by
\bena
\int_{\Omega} \parl{\phi_1}{t} +\frac{\hat \rho_s^{(1)}}{\hat \rho_b^{(1)}}\delta_{\partial \Omega} \parl{\psi_1}{t}d\bx=0, 
\eena
We then introduce the characteristic length $l$, characteristic times $T$ and use the  dimensionless method, we have
\bena
\int_{\hat \Omega} \frac{\hat \rho_b^{(1)}}{\hat \rho_s^{(1)}}l\parl{\phi_1}{\hat t}+\delta_{\partial \hat \Omega} \parl{\psi_1}{\hat t}d\hat \bx=0, 
\eena
where $\hat t=t/T,\hat \Omega=\Omega/{l^d}$,$\partial{\hat \Omega}= \partial \Omega/{l^{d-1}}, \hat \bx=\bx/l^d$ and  $d$ is the number of dimensions.
Introducing $\beta=\frac{\hat \rho_b^{(1)}}{\hat \rho_s^{(1)}}l=\frac{s_i}{v_i}l$ and omitting $\hat{(\cdot)}$, then the equation is reduced to
\bena
\beta\int_\Omega \parl{\phi_1}{t}d\bx + \int_{\partial\Omega} \parl{\psi_1}{t}d\bs=0. \label{volume}
\eena

The equation \eqref{volume} indicates $\beta$ is related to the characteristic length multiplying by the ratio of single molecular volume and effective surface area, which indicates the characteristic lengthscale $l$ determines the magnitude of $\beta$. The equation shows that $\int_{\partial\Omega} \frac{\partial \psi_1}{\partial t} ds$ vanishes as $\beta \rightarrow 0$, meaning mass conservation is required only on the surface when $l \rightarrow 0$. Conversely, $\int_\Omega \frac{\partial \phi_1}{\partial t} d\bx$ becomes zero as $\beta \rightarrow \infty$, indicating that for a very large characteristic length scale, mass conservation in the bulk is dominant. This is consistent with the intuition thinking. It is important to note that in both limiting cases described above, there is no exchange of mass between the bulk and the surface. Consequently, these limiting scenarios should be excluded from discussions specifically focused on bulk-surface material exchange. The above analysis demonstrates that a moderate value of $\beta$ is crucial for PDEs with dynamic boundary conditions to describe bulk-surface interactions, such as material exchange, at the characteristic length scale $l$.

Equation \eqref{volume} shows that a gain of one unit of volume for the first species in the bulk corresponds to a loss of $\beta$ units of area on the surface. Based on this, we then introduce a Robin boundary condition $\alpha f_m=\beta \mu_s-\mu_b$ to describe the exchange of material by connecting the volume outward flux $f_m$, the relaxation parameter $\alpha$, the characteristic length related parameter $\beta$, the chemical potentials $\mu_s$ on surface and $\mu_b$ in bulk. We will introduce the definitions of chemical potentials later in section 2.3. $f_m$ can be viewed as the generalized flux, and $\beta \mu_s-\mu_b$ is the change of energy. The change of energy can also be viewed as the balance of generalized forces.  If $\alpha=0$, then $\beta \mu_s=\mu_b$, which indicates the system reaches the detailed balance of generalized forces, but does not guarantee the equilibrium of phase.  

Since $\Omega$ is arbitrary, from \eqref{volume}, we obtain the equation for the local mass conservation law,
\bena
\beta \parl{\phi_1}{t} + \delta_{\partial \Omega} \parl{\psi_1}{t}=0. \label{volumeevo1}
\eena
Here we introduce two terms $A_b, A_s$, which satisfy
\bena
\parl{\phi_1}{t}=A_b,\\
\parl{\psi_1}{t}=A_s,\\
\beta A_b+\delta_{\partial \Omega} A_s=0. \label{A_b}
\eena
The expressions for $A_b$ and $A_s$ are determined by the second law of thermodynamics and the generalized Onsager principle, as detailed in Section 2.3.
\begin{remark}
	If the order parameters are chosen as the mass fractions $\frac{\rho_b^{(i)}}{\rho_b^{(1)}+\rho_b^{(2)}}$ and $\frac{\rho_s^{(i)}}{\rho_s^{(1)}+\rho_s^{(2)}}$, the dynamic is trivial and meaningless.
\end{remark}
\subsection{Volume conservation law with the exchange of material  between the bulk and surface} 
\noindent \indent Secondly, we can also discuss it from the viewpoint of volume conservation in polymer system. We define $n_b^{(i)}$ and $n_s^{(i)}$ as the number density of the $i^{th}$ species of polymer in the bulk and on the surface, and we assume total number of the $i^{th}$ species of polymer is a constant, which is
\bena
\frac{d}{dt}[\int_\Omega n_b^{(i)}d\bx+\int_{\partial \Omega} n_s^{(i)}d\bs ]=0.
\eena
We then introduce the volume fraction $\phi_i=\frac{n_b^{(i)}N_i}{\Theta_b}$ and area fraction $\psi_i=\frac{n_s^{(i)}N_i}{\Theta_s}$, where $N_i$ is the degree of polymerization (the number of monomer in a polymer chain), $\Theta_b=\sum_i n_b^{(i)}N_i$ and $\Theta_s=\sum_i n_s^{(i)}N_i$ are two constant number densities of monomers when introducing the assumption of volume exclusion. We then have 
\bena
\frac{\phi_i}{\psi_i}=\frac{n_b^{(i)}\Theta_s}{n_s^{(i)}\Theta_b},
\eena
and \bena
\frac{d}{dt}[\frac{\Theta_b}{N_i}\int_\Omega\phi_i d\bx+\frac{\Theta_s}{N_i}\int_{\partial \Omega} \psi_i d\bs ]=0.
\eena
Introducing the characteristic length $l$ and characteristic time  $T$, by using the dimensionless method, we have
\bena
\frac{d}{d\hat t}[\frac{\Theta_bl}{\Theta_s}\int_{\hat \Omega}\phi_i d \hat \bx+\int_{\partial  {\hat \Omega}} \psi_i d \hat \bs ]=0,
\eena
where $\hat t=t/T, \hat \bx=x/l^d, \hat \bs=\bs/l^{d-1}, \hat \Omega =\Omega/l^d,\partial{\hat \Omega} =\partial{\Omega/l^{d-1}},$ and $d$ is the dimension number.

By introducing $\beta=\frac{\Theta_bl}{\Theta_s}$ and omitting $\hat{(\cdot)}$, then the physical meaning of $\beta$ is the characteristic length multiplying the ratio of carrying capacity of monomers between the bulk and surface. Since $\Omega$ is arbitrary, so the local volume conservation equation is given by
\bena
\beta \parl{\phi_i}{t} + \delta_{\partial \Omega} \parl{\psi_i}{t}=0.
\eena
 In this section, the physical meaning of $\beta$ is analogous to that in section 2.1. There is no essential difference of $\beta$ in section 2.1 and 2.2. We then discuss the exchange of energy and energy dissipation based on such general dynamic system. Without loss the generality, we set $\phi=\phi_1$ and $\psi=\psi_1$.
\subsection{Energy dissipation law with the exchange of energy between bulk and surface}
 \noindent \indent Now let us develop the thermodynamically consistent phase field model with dynamic boundary conditions. For the small system with finite size, the free energy should be modified by adding the effects of the boundary. As we known, the exchange of energy between bulk and surface would be significant. We firstly introduce the nondimensional free energy of derivatives up to the second order in space in the bulk,
\ben
E_b[\phi]=\int_{\Omega} e_b( \phi, \phi_t,\nabla \phi, \nabla \nabla \phi) d\bx,
\een
where $e_b$ is the energy density per unit volume and $\phi_t$ meansures the inertia effects of the system. We consider a binary  material system with a boundary that may have its distinctive properties than the bulk and possesses its own nondimensional surface energy and coupling energy accounting for the free energy due to the discrepancy between the two physical quantities
\bena
E_s[\psi]=\int_{\partial \Omega} e_s( \psi, \psi_t,\nabla_s \psi, \nabla_s \nabla_s \psi) d\bs,\\
E_c[\phi, \psi]=\int_{\partial \Omega} e_c(\phi, \nabla_s \phi, \nabla_s \nabla_s \phi,\psi, \nabla_s \psi, \nabla_s \nabla_s \psi) d\bs,
\eena
where $\psi(\bs, t)$ is not necessarily the same as $\phi(\bs, t)=\lim_{\bx \rightarrow \bs}\phi(\bx,t)$ confined to the surface as we discussed in section 2.1,   $e_s, e_c$ are the surface energy density per unit, and $\psi_t$ is used to measure the inertia effects on the surface
and $\nabla_s$ is the surface gradient operator over  smooth boundary $\partial \Omega$ as those in \cite{dziuk2013finite,brenner2013interfacial}.
The total nondimensional free energy of the system is given by
\bena
E[\phi,\psi]=\int_{\Omega} e_b(\phi, \phi_t,\nabla \phi, \nabla \nabla \phi) d\bx+\int_{\partial \Omega} e_s(\psi, \psi_t,\nabla_s \psi, \nabla_s \nabla_s \psi)+e_c(\phi, \nabla_s \phi, \nabla_s \nabla_s \phi,\psi, \nabla_s \psi, \nabla_s \nabla_s \psi)d\bs.
\eena

We calculate the  time rate of change of the free energy as follows, assuming  domain $\Omega$ is fixed,
	\ben
	\bea{l}
\frac{dE}{dt}=\int_{\Omega} \mu_b  \phi_t d\bx+\int_{\partial \Omega} [\frac{\partial e_s}{\partial \psi_t} \psi_{tt}+\frac{\partial (e_s+e_c)}{\partial \psi} \psi_t+\frac{\partial (e_s+e_c)}{\partial  \nabla_s \psi} \nabla_s \psi_t +\frac{\partial (e_s+e_c)}{\partial  \nabla_s \nabla_s \psi} \nabla_s \nabla_s \psi_t\\
\qquad +\frac{\partial (e_s+e_c)}{\partial \phi} \phi_t+\frac{\partial (e_s+e_c)}{\partial  \nabla_s \phi} \nabla_s \phi_t +\frac{\partial (e_s+e_c)}{\partial  \nabla_s \nabla_s \phi} \nabla_s \nabla_s \phi_t+
\bn \cdot \frac{\partial e_b}{\partial \nabla \phi} \phi_t+\frac{\partial e_b}{\partial \nabla \nabla \phi}: \bn \nabla \phi_t-\bn \nabla: \frac{\partial e_b}{\partial \nabla \nabla \phi} \phi_t ] d\bs\\
=\int_{\Omega} \mu_b  \phi_t d\bx+\int_{\partial \Omega} [\rho_s \psi_t \psi_{tt}+\frac{\partial (e_s+e_c)}{\partial \psi} \psi_t-\nabla_s \cdot \frac{\partial (e_s+e_c)}{\partial  \nabla_s \psi} \psi_t-2H \bn \cdot \frac{\partial (e_s+e_c)}{\partial  \nabla_s \psi} \psi_t -\nabla_s \cdot \frac{\partial (e_s+e_c)}{\partial  \nabla_s \nabla_s \psi} \cdot \nabla_s \psi_t\\
\qquad-2H \bn \cdot \frac{\partial ( e_s+e_c)}{\partial \nabla_s  \nabla_s \psi} \cdot \nabla_s \psi_t+
\frac{\partial (e_s+e_c)}{\partial \phi} \phi_t-\nabla_s \cdot \frac{\partial (e_s+e_c)}{\partial  \nabla_s \phi} \phi_t-2H \bn \cdot \frac{\partial (e_s+e_c)}{\partial  \nabla_s \phi} \phi_t -\nabla_s \cdot \frac{\partial (e_s+e_c)}{\partial  \nabla_s \nabla_s \phi} \cdot \nabla_s \phi_t\\
\qquad-2H \bn \cdot \frac{\partial ( e_s+e_c)}{\partial \nabla_s  \nabla_s \phi} \cdot \nabla_s \phi_t+
\bn \cdot \frac{\partial e_b}{\partial \nabla \phi} \phi_t +\frac{\partial e_b}{\partial \nabla \nabla \phi}: \bn \nabla \phi_t-(\bn \nabla: \frac{\partial e_b}{\partial \nabla \nabla \phi}) \phi_t ] d\bs\\
=\int_{\Omega} \mu_b \phi_t d\bx+\int_{\partial \Omega} [\mu_s  \psi_t+\mu_c \phi_t+\mu_g (\bn \cdot \nabla \phi_t ) ] d\bs,
	\eea\label{disp-eq}
	\een

where we adopt the Einstein notation for tensors here, denote tensor product of vector $\bn$ and $\bv$ as $\bn\bv=n_i v_j$,  use one dot $\cdot$ to represent inner product $\bn\cdot \bv=n_i v_i$ and two dots $:$ to represent contraction of two second order tensor $\bA:\bC=A_{ij} C_{ij}$, where $\bn,\bv$ are vectors, $\bA, \bC$ are second order tensors, $H=\frac{1}{2}\nabla_s \cdot \bn$ is the mean curvature of the boundary, $\bn$ is the unit external normal of $\partial \Omega$, the~bulk chemical potential $\mu_b$, surface chemical potentials $\mu_s$, $\mu_c$, and the conjugate variable $\mu_g$ are given, respectively,  by
	\bena
	\mu_b=\rho \phi_{tt}+\frac{\partial e_b}{\partial \phi}-\nabla \cdot \frac{\partial e_b}{\partial \nabla \phi}+\nabla \nabla : \frac{\partial e_b}{\partial \nabla \nabla \phi},\\
	\mu_s=\rho_s \psi_{tt}+\frac{\partial (e_s+e_c)}{\partial \psi}-\nabla_s \cdot \frac{\partial (e_s+e_c)}{\partial  \nabla_s \psi}-2H \bn \cdot \frac{\partial (e_s+e_c)}{\partial  \nabla_s \psi}+\nabla_s \nabla_s : \frac{\partial (e_s+e_c)}{\partial  \nabla_s \nabla_s \psi}+2H \bn \nabla_s: \frac{\partial (e_s+e_c)}{\partial \nabla_s  \nabla_s \psi} +\\
	\qquad \nabla_s \cdot (2H\bn \cdot \frac{\partial (e_s+e_c)}{\partial \nabla_s  \nabla_s \psi})+4H^2\bn \bn :\frac{\partial (e_s+e_c)}{\partial \nabla_s  \nabla_s \psi},\\
	\mu_c=\frac{\partial e_c}{\partial \phi}-\nabla_s \cdot \frac{\partial e_c}{\partial  \nabla_s \phi}-2H \bn \cdot \frac{\partial e_c}{\partial  \nabla_s \phi}+\nabla_s \nabla_s : \frac{\partial e_c}{\partial  \nabla_s \nabla_s \phi}+2H \bn \nabla_s: \frac{\partial e_c}{\partial \nabla_s  \nabla_s \phi} +\\
	\qquad \nabla_s \cdot (2H\bn \cdot \frac{\partial e_c}{\partial \nabla_s  \nabla_s \phi})+4H^2\bn \bn :\frac{\partial e_c}{\partial \nabla_s  \nabla_s \phi}+\bn \cdot \frac{\partial e_b}{\partial \nabla \phi}-\bn \nabla : \frac{\partial e_b}{\nabla \nabla \phi}-\nabla_s\bn:\frac{\partial e_b}{\partial \nabla \nabla \phi} -2H\bn \bn :\frac{\partial e_b}{\partial \nabla \nabla \phi},\\
	\mu_g=\frac{\partial e_b}{\partial \nabla \nabla \phi}: \bn \bn.
		\eena
		\begin{remark}
 (i) There exist two surface chemical potentials, one corresponding to $\phi$ is denoted as $\mu_c$, while the other corresponding to $\psi$ is denoted as $\mu_s$ at the boundary. The~surface chemical potential, $\mu_c$, includes contributions from the bulk energy confined to the boundary, while $\mu_s$ only includes contributions from the surface.  (ii)
	The mean curvature shows up in the surface chemical potentials, indicating that curvature of the boundary affects the surface dynamics. (iii) More surface terms can appear if the free energy density function depends on higher order spatial derivatives. 
\end{remark}
From the second law of thermodynamics and statistical mechanics, the Helmholtz free energy should be dissipated with the constant temperature and fixed domain.	In the following we will discuss the dynamics of exchange of material and energy  by using the generalized Onsager principle. When there is no exchange of material and energy by setting $\bn \cdot \nabla \mu_b=0$, the boundary condition can also change with time. However, this is a special case for the model with dynamic boundary conditions, we have proved this by using asymptotic analysis method in our previous work, thus we will not repeat it in this paper.

\section{Generalized Onsager principle, dynamics and energy dissipation}
\subsection{Generalized Onsager principle}
 \noindent \indent The generalized Onsager principle (GOP) is build on the Onsager  reciprocal relation and linear response theorem, it can be simplified for the phase field model and summarized as following,
\bena
\bP=-\bM \bQ,
\eena
where $\bP$ is a vector respecting the generalized fluxes, $\bM$ is the semi-positive definite moblity operator and can be always reformulated by the summation of a sysmetric mobility $\bM^{s}$ and a anti-sysmetric mobility $\bM^{a}$, $\bQ$ is a vector respecting the generlized forces, which is also called as the chemical potentials.
\subsection{Dynamics in the bulk}	
 \noindent \indent 	We apply the generalized Onsager principle firstly to the bulk integral in \eqref{disp-eq} to obtain the dynamic equation for $\phi$ in the bulk $\Omega$,
	\ben
	\phi_t=-M_b\mu_b, \quad \bx \in \Omega, \label{bulk-eq}
	\een
	where $M_b$ is the  semi-positive definite mobility operator to ensure energy dissipation.  We firstly introduce the mobility operator in the following form in this study
	\ben
	M_b=M_b^{(1)}-\nabla \cdot \bM_b^{(2)} \cdot \nabla,
	\een
	where $M_b^{(1)}$ is a scalar function of $\phi$, $ \bM_b^{(2)} \in \bR^{3\times 3}$ is a semi-definite positive matrix which can be a function of $\phi$ as well. If~$ \bM_b^{(2)}=M_b^{(2)}\bI$ and $M_b^{(2)}$ is a scalar function of $\phi$, then a special case $\nabla \cdot \bM_b^{(2)} \cdot \nabla=\nabla \cdot (M_b^{(2)} \nabla)$ can be obtained. 
	With this, the~energy dissipation rate   reduces to

		\ben
		\bea{l}
		\frac{dE}{dt}=- \int_{\Omega} [\mu_b M_b^{(1)}\mu_b+\nabla \mu_b  \cdot \bM_b^{(2)} \cdot \nabla \mu_b] d\bx+
		\int_{\partial \Omega} [\mu_s\psi_t+\mu_c\phi_t+\mu_g \nabla_\bn \phi_t+\mu_b \bn \cdot \bM_b^{(2)} \cdot \nabla \mu_b] ds. \label{energy-diss}
		\eea
		\een
	We remark that $f_m=\bn \cdot \bM_b^{(2)} \cdot \nabla \mu_b$ is the outward flux
	 across the boundary. This  physical quantity is determined by the balance between the surface potential  $\mu_s$ and bulk chemical potential  $\mu_b$ as discussion in section 2.1. We next derive thermodynamically  consistent boundary conditions  for this~model.

	\subsection{Dynamics on the boundary}
	
	\noindent \indent We then apply the generalized Onsager principle the second time to the surface generalized fluxes and forces to establish a dynamic constitutive equation at the boundary:
	\ben
	\left (
	\bea{l}
	\phi_t\\
	\psi_t\\
	f_m\\
	\nabla_{\bn}\phi_t
	\eea
	\right)
	=-\bM_{4\times 4} \cdot
	\left (
	\bea{l}
	\mu_c\\
	\mu_s\\
	\mu_b\\
	\mu_g
	\eea
	\right),\label{GOP-s}
	\een
	where $f_m=\bn \cdot  \bM_b^{(2)} \cdot \nabla\mu_b $ is the outward volume flux, $\phi_t, \psi_t, \nabla_\bn \phi_t$ are time rate of changes of three quantities, identified as generalized fluxes,  and~$\bM_{4\times 4}\geq 0$ is the surface mobility operator, a~$4\times 4$ matrix or second-order tensor. $\bM_{4\times4}\geq0$ means that its symmetric part is semi-positive definite definite, which guarantees the system satisfies the second law of thermodynamics.
	
	Based on the discussion in section 2, we specify the mobility operator of surface as follows
	\ben
	\bM_{4\times 4}=
	\left (
	\bea{lcrcr}
	M_{11}  &M_{12} &M_{13}&M_{14}\\
	M_{21} & M_{22} & M_{23} & M_{24}\\
	0 & -\frac{\beta}{\alpha} & \frac{1}{\alpha} & 0
	\\
	M_{41}& M_{42} & M_{43} & M_{44}
	\eea
	\right).\label{purely}
	\een
	where we used the relationship $\alpha f_m=\beta \mu_s-\mu_b$, $\alpha > 0$ is a friction coefficient to control the relaxation,  $\beta $ is weight coefficients related to the characteristic length, which is related by the system as we discussed in section 2. 
%
	
According to the Generalized Onsager Principle (GOP), the mobility operator $\mathbf{M}_{4\times4}$ can be decomposed as $\mathbf{M}_{4\times4} = \mathbf{M}_{4\times4}^{(s)} + \mathbf{M}_{4\times4}^{(a)}$, where $\mathbf{M}_{4\times4}^{(s)}$ is symmetric and accounts for irreversible processes, while $\mathbf{M}_{4\times4}^{(a)}$ is antisymmetric and corresponds to reversible processes.

We now proceed to formulate the surface mobility operator by considering two types of transport processes: the first describes a purely irreversible transport system (Model A), and the second incorporates both irreversible and reversible dynamics (Model B). Furthermore, for reactive transport systems, we develop Model C and Model D to represent the purely irreversible and irreversible-reversible cases, respectively.
\section{Thermodynamically consistent phase filed models with dynamic boundary conditions}
\noindent \indent This section derives the governing equations and corresponding dynamic boundary conditions for systems involving both irreversible and reversible processes, encompassing pure transport as well as reactive transport.. 
\subsection{Model A for irreversible transport system}
\noindent \indent For the pure irreversible transport system,  $\bM_{4\times4}=\bM_{4\times4}^{(s)}$, then 
	\ben
\bM_{4\times 4}=
\left (
\bea{lcrcr}
M_{11}  &M_{12} &0 &M_{14}\\
M_{12} & M_{22} & -\frac{\beta}{\alpha}  & M_{24}\\
0 & -\frac{\beta}{\alpha} & \frac{1}{\alpha} & 0
\\
M_{14}& M_{24} & 0 & M_{44}
\eea
	\right).\label{purely}
\een
To check the conservation law of mass, then
	\bena
0=\beta \int_\Omega \parl{\phi}{t}d\bx+\int_{\partial \Omega} \parl {\psi}{t}d\bs
=\beta \int_\Omega-M_b^{(1)}\mu_bd\bx+\int_{\partial \Omega}-M_{12}\mu_c+(\frac{\beta^2}{\alpha}-M_{22})\mu_s-M_{24}\mu_gd\bs,
\eena
The conservation law will introduce constraints for the mobility operator. It is obviously, $M_b^{(1)}$ should be zero, the other mobility operators should be  prudently chosen to satisfy the conservation law of mass. Now we set $M_b^{(1)}=0,M_{12}=-\nabla_s\cdot \bM_{12}^{(2)} \cdot \nabla_s,M_{22}=\frac{\beta^2}{\alpha}-\nabla_s\cdot\bM_{22}^{(2)}\cdot \nabla_s,M_{24}=-\nabla_s\cdot \bM_{24}^{(2)}\cdot\nabla_s,$ we can get
	\bena
\beta \int_\Omega \parl{\phi}{t}d\bx+\int_{\partial \Omega} \parl {\psi}{t}d\bs
=-2H \int_{\partial \Omega}\bn \cdot \bM_{12}^{(2)} \cdot \nabla_s \mu_c-\bn \cdot\bM_{22}^{(2)}\cdot \nabla_s \mu_s+\bn \cdot \bM_{24}^{(2)} \cdot \nabla_s \mu_gd\bs.
\eena
If $H=0$ or $\bM_{12}^{(2)}, \bM_{22}^{(2)}, \bM_{24}^{(2)}$are scalars multiplied by unit matrix, then $\beta \int_\Omega \parl{\phi}{t}d\bx+\int_{\partial \Omega} \parl {\psi}{t}d\bs=0$. If $H=0$, the mobility of the surface is given by
\ben
\bM_{4\times 4}=
\left (
\bea{lcrcr}
M_{11}  &-\nabla_s\cdot \bM_{12}^{(2)}\cdot \nabla_s &0 &M_{14}\\
-\nabla_s\cdot \bM_{12}^{(2)}\cdot \nabla_s  & \frac{\beta^2}{\alpha}-\nabla_s\cdot \bM_{22}^{(2)}\cdot \nabla_s& -\frac{\beta}{\alpha}  & -\nabla_s\cdot \bM_{24}^{(2)} \cdot \nabla_s\\
0 & -\frac{\beta}{\alpha} & \frac{1}{\alpha} & 0
\\
M_{14}& -\nabla_s\cdot \bM_{24}^{(2)} \cdot \nabla_s & 0 & M_{44}
\eea
\right).
\een
 For the general case, $H\neq 0$, so the mobility operator on the surface should be
	\ben
\bM_{4\times 4}=
\left (
\bea{lcrcr}
M_{11}  &-\nabla_s\cdot (M_{12}^{(2)}\nabla_s) &0 &M_{14}\\
-\nabla_s\cdot (M_{12}^{(2)}\nabla_s) & \frac{\beta^2}{\alpha}-\nabla_s\cdot (M_{22}^{(2)}\nabla_s) & -\frac{\beta}{\alpha}  & -\nabla_s\cdot (M_{24}^{(2)}\nabla_s)\\
0 & -\frac{\beta}{\alpha} & \frac{1}{\alpha} & 0
\\
M_{14}& -\nabla_s\cdot (M_{24}^{(2)}\nabla_s) & 0 & M_{44}
\eea
\right). \label{pure1}
\een

By setting $M_{11}=M_c, M_{14}=0, M_{44}=M_g$ and $M_{24}=0$ for simplicity, the governing system of equations together with the dynamic boundary conditions is summarized as follows.
\bena\label{g-model1}
\phi_t=\nabla \cdot \bM_b^{(2)} \cdot \nabla \mu_b,  \quad \bx \in \Omega, \\
\phi_t=-M_c \mu_c+\nabla_s\cdot (M_{12}^{(2)}\nabla_s \mu_s), \quad \bs \in \partial \Omega, \\
\psi_t=\nabla_s\cdot (M_{12}^{(2)}\nabla_s \mu_c)-[\frac{\beta^2}{\alpha}-\nabla_s\cdot (M_{22}^{(2)}\nabla_s)]\mu_s+\frac{\beta}{\alpha}\mu_b, \quad \bs \in \partial \Omega, \\
\alpha \bn \cdot \bM_b^{(2)} \cdot \nabla \mu_b=-\mu_b+\beta \mu_s, \quad
\nabla_\bn\phi_t=-M_g \mu_g,  \quad \bs \in \partial \Omega, 
\eena
so we can get $A_b=\nabla \cdot \bM_b^{(2)} \cdot \nabla \mu_b, A_s=\nabla_s\cdot (M_{12}^{(2)}\nabla_s \mu_c)-[\frac{\beta^2}{\alpha}-\nabla_s\cdot (M_{22}^{(2)}\nabla_s)]\mu_s+\frac{\beta}{\alpha}\mu_b$ in \eqref{A_b}.
The corresponding  energy dissipation rate is given by
\ben
\bea{l}
\frac{dE}{dt}=- \int_{\Omega} \nabla \mu_b  \cdot \bM_b^{(2)} \cdot \nabla \mu_b d\bx-
\int_{\partial \Omega} \bU \cdot \bM_{4\times 4} \cdot \bU d\bs\leq 0,\label{energy-diss1}
\eea
\een
where
$
\bU=
\left (
\bea{lcrcr}
\phi_t,&
\psi_t,&
f_m,&
\nabla_{\bn}\phi_t
\eea
\right)^T
$, $\bM_{4\times 4}$ is semi-positive definite.

\begin{remark}
	Both the second law of thermodynamics and mass conservation law are two constraints for the moblity operators. The mobility in \eqref{purely} is only constrained by the second law of thermodynamics, however, the mobility in \eqref{pure1} is constrained by the both laws.  
\end{remark}
\begin{remark}
In our previous work in \cite{jing2022thermodynamically}, we set
\ben
\bM_{4\times 4}=
\left (
\bea{lcrcr}
M_c  &0 &0&0\\
0 & M_s+\frac{\beta^2}{\alpha} & -\frac{\beta}{\alpha} & 0\\
0 & -\frac{\beta}{\alpha} & \frac{1}{\alpha} & 0
\\
0 & 0 & 0 & M_g
\eea
\right)
\een
which is the special case of \eqref{pure1} by setting $M_{11}=M_c$, $M_{12}=M_{14}=M_{24}=0$. 
\end{remark}

\subsection{Model B for irreversible-reversible transport system} 
\noindent \indent For a system including both irreversible and reversible processes, $\bM_{4\times4}=\bM_{4\times4}^{(s)}+\bM_{4\times4}^{(a)}$, the moblity operator can be reformulated to 
\bena
\bM_{4\times 4}=
\left (
\bea{lcrcr}
M_{11}^{(s)}  &M_{12}^{(s)} &M_{13}^{(s)}&M_{14}^{(s)}\\
M_{12}^{(s)} & M_{22}^{(s)} & M_{23}^{(s)}  & M_{24}^{(s)}\\
M_{13}^{(s)} & M_{23}^{(s)} & \frac{1}{\alpha} & M_{34}^{(s)}
\\
M_{14}^{(s)}& M_{24}^{(s)} & M_{34}^{(s)} & M_{44}^{(s)}
\eea
\right)+\left (
\bea{lcrcr}
0  &M_{12}^{(a)} &M_{13}^{(s)} &M_{14}^{(a)}\\
-M_{12}^{(a)} & 0 & \frac{\beta}{\alpha}+M_{23}^{(s)}  & M_{24}^{(a)}\\
-M_{13}^{(s)} & -\frac{\beta}{\alpha}-M_{23}^{(s)} & 0 & -M_{34}^{(s)}
\\
-M_{14}^{(a)}& -M_{24}^{(a)} & M_{34}^{(s)} & 0
\eea
\right)
\\
\qquad =\left (
\bea{lcrcr}
M_{11}^{(s)}  &M_{12}^{(s)}+M_{12}^{(a)} &2M_{13}^{(s)} &M_{14}^{(s)}+M_{14}^{(a)}\\
M_{12}^{(s)}-M_{12}^{(a)} & M_{22}^{(s)} & \frac{\beta}{\alpha}+2M_{23}^{(s)}  & M_{24}^{(s)}+M_{24}^{(a)}\\
0 & -\frac{\beta}{\alpha} & \frac{1}{\alpha} & 0
\\
M_{14}^{(s)}-M_{14}^{(a)}& M_{24}^{(s)}-M_{24}^{(a)} & 2M_{34}^{(s)} & M_{44}^{(s)}
\eea
\right),\label{non-purely1}
\eena
where the symmetric and anti-symmetric mobility contributes to the irreversible and reversible transport dynamics at the boundary, respectively. The~antisymmetric mobility component represents an energy exchange between the bulk and the boundary without inducing any energy dissipation.

To check the conservation law of mass, using \eqref{non-purely1}, then
\bena
0=\beta \int_\Omega \parl{\phi}{t}d\bx+\int_{\partial \Omega} \parl {\psi}{t}d\bs\\
=\beta \int_\Omega-M_b^{(1)}\mu_bd\bx+\int_{\partial \Omega}-(M_{12}^{(s)}-M_{12}^{(a)})\mu_c+(\frac{\beta^2}{\alpha}-M_{22}^{(s)})\mu_s-2(\frac{\beta}{\alpha}+M_{23}^{(s)})\mu_b-(M_{24}^{(s)}+M_{24}^{(a)})\mu_gd\bs.
\eena
To satisfy the conservation law, not set $M_b^{(1)}=0, M_{12}^{(s)}-M_{12}^{(a)}=-\nabla_s \cdot \bM_{12}^{(2)}\cdot \nabla_s,M_{22}^{(s)}=\frac{\beta^2}{\alpha}-\nabla_s \cdot \bM_{22}^{(2)}\cdot \nabla_s,M_{23}^{(s)}=-\nabla_s \cdot \bM_{23}^{(2)}\cdot \nabla_s-\frac{\beta}{\alpha},M_{24}^{(s)}=-M_{24}^{(a)}-\nabla_s \cdot \bM_{24}^{(2)} \cdot \nabla_s$,
then 
\bena
0=\beta \int_\Omega \parl{\phi}{t}d\bx+\int_{\partial \Omega} \parl {\psi}{t}d\bs\\
=-2H \int_{\partial \Omega} \bn\cdot \bM_{12}^{(2)}\cdot \nabla_s\mu_c+\bn \cdot \bM_{22}^{(2)}\cdot \nabla_s \mu_s+2\bn \cdot \bM_{23}^{(2)}\cdot \nabla_s \mu_b+\bn \cdot \bM_{24}^{(2)}\cdot \nabla_s\mu_gd\bs.
\eena
If $H=0$ or $\bM_{12}^{(2)},\bM_{22}^{(2)},\bM_{23}^{(2)},\bM_{24}^{(2)}$ are scalars multiplied by unit matrix, then $\beta \int_\Omega \parl{\phi}{t}d\bx+\int_{\partial \Omega} \parl {\psi}{t}d\bs=0$. $H=0$ is the simplest case, we omit the discussion of the mobility on the surface in this case. When $H \neq 0$ and $\bM_{12}^{(2)},\bM_{22}^{(2)},\bM_{23}^{(2)},\bM_{24}^{(2)}$ are scalars multiplied by unit matrix,  the mobility operator on the surface is given by
\bena
\bM_{4\times 4} =\left (
\bea{lcrcr}
M_{11}^{(s)}  &2M_{12}^{(a)}-\nabla_s \cdot (M_{12}^{(2)}\nabla_s) &2M_{13}^{(s)} &M_{14}^{(s)}+M_{14}^{(a)}\\
-\nabla_s \cdot (M_{12}^{(2)}\nabla_s) & \frac{\beta^2}{\alpha}-\nabla_s \cdot (M_{22}^{(2)}\nabla_s) & -\frac{\beta}{\alpha}-2\nabla_s \cdot (M_{23}^{(2)}\nabla_s)  & -\nabla_s \cdot (M_{24}^{(2)}\nabla_s)\\
0 & -\frac{\beta}{\alpha} & \frac{1}{\alpha} & 0
\\
M_{14}^{(s)}-M_{14}^{(a)}& -2M_{24}^{(a)}-\nabla_s \cdot (M_{24}^{(2)}\nabla_s) & 2M_{34}^{(s)}& M_{44}^{(s)}
\eea
\right)
=\bM_{4\times 4}^{(a)}+\bM_{4\times 4}^{(s)},\\
\bM_{4\times 4}^{(a)}=\left (
\bea{lcrcr}
0  &M_{12}^{(a)} &M_{13}^{(s)} &M_{14}^{(a)}\\
-M_{12}^{(a)} & 0 & -\nabla_s \cdot (M_{23}^{(2)}\nabla_s)   & M_{24}^{(a)}\\
-M_{13}^{(s)} & \nabla_s \cdot (M_{23}^{(2)}\nabla_s)  & 0 & -M_{34}^{(s)}
\\
-M_{14}^{(a)}& -M_{24}^{(a)} & M_{34}^{(s)} & 0
\eea
\right)\\
\bM_{4\times 4}^{(s)}=\left (
\bea{lcrcr}
M_{11}^{(s)}  &M_{12}^{(a)}-\nabla_s \cdot (M_{12}^{(2)}\nabla_s)  &M_{13}^{(s)}&M_{14}^{(s)}\\
M_{12}^{(a)}-\nabla_s \cdot (M_{12}^{(2)}\nabla_s)  & \frac{\beta^2}{\alpha}-\nabla_s \cdot (M_{22}^{(2)}\nabla_s) & -\frac{\beta}{\alpha}-\nabla_s \cdot (M_{23}^{(2)}\nabla_s)  & -M_{24}^{(a)}-\nabla_s \cdot (M_{24}^{(2)}\nabla_s)\\
M_{13}^{(s)} & -\frac{\beta}{\alpha}-\nabla_s \cdot (M_{23}^{(2)}\nabla_s) & \frac{1}{\alpha} & M_{34}^{(s)}
\\
M_{14}^{(s)}& -M_{24}^{(a)}-\nabla_s \cdot (M_{24}^{(2)}\nabla_s) & M_{34}^{(s)} & M_{44}^{(s)}
\eea
\right)
.\label{non-purely2}
\eena 
\begin{remark}
	There are connections between the symmetric mobility operator $\bM_{4\times 4}^{(s)} $ and anti-symmetric mobility operator $\bM_{4\times 4}^{(a)} $ by sharing some common elements like $M_{13}^{(s)}$ and $\nabla_s \cdot (M_{23}^{(2)}\nabla_s)$. 
\end{remark}
To clarify the mechanism of exchange of material and energy between bulk and surface, we simplify the mobility operator on the surface by setting $M_{12}^{(a)}=M_{14}^{(s)}=M_{14}^{(a)}=M_{24}^{(a)}=M_{24}^{(2)}=M_{34}^{(s)}=0, M_{11}^{(s)}=M_c,M_{44}^{(s)}=M_g$, then
\bena
\bM_{4\times 4} =\left (
\bea{lcrcr}
M_{c}  &-\nabla_s \cdot (M_{12}^{(2)}\nabla_s) &2M_{13}^{(s)} &0\\
-\nabla_s \cdot (M_{12}^{(2)}\nabla_s) & \frac{\beta^2}{\alpha}-\nabla_s \cdot (M_{22}^{(2)}\nabla_s) & -\frac{\beta}{\alpha}-2\nabla_s \cdot (M_{23}^{(2)}\nabla_s)  & 0\\
0 & -\frac{\beta}{\alpha} & \frac{1}{\alpha} & 0
\\
0& 0 & 0 & M_{g}
\eea
\right). \label{non-purely3}
\eena
The governing system of equations together with the dynamic boundary conditions in this model is summarized as follows.

\bena\label{g-model2}
\phi_t=\nabla \cdot \bM_b^{(2)} \cdot \nabla \mu_b,  \quad \bx \in \Omega, \\
\phi_t=-M_c \mu_c+\nabla_s \cdot (M_{12}^{(2)}\nabla_s \mu_s)+2M_{13}^{(s)}\mu_b,\quad \bs \in \partial \Omega, \\
\psi_t=\nabla_s \cdot (M_{12}^{(2)}\nabla_s \mu_c)-[\frac{\beta^2}{\alpha}-\nabla_s \cdot (M_{22}^{(2)}\nabla_s)]\mu_s+[\frac{\beta}{\alpha}+2\nabla_s \cdot (M_{23}^{(2)}\nabla_s)] \mu_b, \quad \bs \in \partial \Omega, \\
\alpha \bn \cdot \bM_b^{(2)} \cdot \nabla \mu_b=-\mu_b+\beta \mu_s, \quad
\nabla_\bn\phi_t=-M_g \mu_g,  \quad \bs \in \partial \Omega. 
\eena

The corresponding  energy dissipation rate is given by
\ben
\bea{l}
\frac{dE}{dt}=- \int_{\Omega} \nabla \mu_b  \cdot \bM_b^{(2)} \cdot \nabla \mu_b d\bx-
\int_{\partial \Omega} \bU \cdot \bM_{4\times 4} \cdot \bU d\bs\\
=- \int_{\Omega} \nabla \mu_b  \cdot \bM_b^{(2)} \cdot \nabla \mu_b d\bx-
\int_{\partial \Omega} \bU \cdot \bM_{4\times 4}^{(s)}\cdot \bU d \bs\leq 0,\label{energy-diss2}
\eea
\een
where
$
\bU=
	\left (
\bea{lcrcr}
\phi_t,&
\psi_t,&
f_m,&
\nabla_{\bn}\phi_t
\eea
\right)^T
$, $\bM_{4\times 4}^{(s)}$ is semi-positive definite,  $\bM_{4\times 4}^{(a)}$ does not contribute to the energy dissipation rate but affect the exchange of material between the bulk and surface.

\begin{remark}
\eqref{g-model1} and \eqref{g-model2} are two different dynamic boundary conditions, the significant difference between them are the time evolution of $\psi$ on the surface besides the energy dissipation rates. In \eqref{g-model1},
\bena
\phi_t=-M_c \mu_c+\nabla_s \cdot (M_{12}^{(2)}\nabla_s \mu_s),\quad \bs \in \partial \Omega, \\
\psi_t=\nabla_s \cdot (M_{12}^{(2)}\nabla_s^2 \mu_c)-[\frac{\beta^2}{\alpha}-\nabla_s \cdot (M_{22}^{(2)}\nabla_s)]\mu_s+\frac{\beta}{\alpha}\mu_b\\
\quad=\nabla_s \cdot (M_{12}^{(2)}\nabla_s \mu_c)+\nabla_s \cdot (M_{22}^{(2)}\nabla_s\mu_s)-\beta f_m,\quad \bs \in \partial \Omega.
\eena
In \eqref{g-model2},
\bena
\phi_t=-M_c \mu_c+\nabla_s \cdot (M_{12}^{(2)}\nabla_s \mu_s)+2M_{13}^{(s)}\mu_b,\quad \bs \in \partial \Omega, \\
\psi_t=\nabla_s \cdot (M_{12}^{(2)}\nabla_s \mu_c)-[\frac{\beta^2}{\alpha}-\nabla_s \cdot (M_{22}^{(2)}\nabla_s)]\mu_s+[\frac{\beta}{\alpha}+2\nabla_s \cdot (M_{23}^{(2)}\nabla_s)] \mu_b\\
\quad=\nabla_s \cdot (M_{12}^{(2)}\nabla_s \mu_c)+\nabla_s \cdot (M_{22}^{(2)}\nabla_s\mu_s)+2\nabla_s \cdot (M_{23}^{(2)}\nabla_s \mu_b)-\beta f_m, \quad \bs \in \partial \Omega.
\eena
If $M_{23}^{(2)}=M_{13}^{(s)}=0$, the above two dynamics are same, so the first model is a special case of the second model by setting $\bM_{4\times 4}^{(a)}=0$ mathematically. $M_{23}^{(2)}\neq 0$ indicates the reversible exchanges of material and energy are controlled by the bulk chemical potential $\mu_b$ and mobility operators $\nabla_s \cdot (M_{23}^{(2)}\nabla_s)$. Since $\alpha >0$, if $\beta \mu_s >\mu_b$, so $f_m>0$, then  $-\beta f_m$ is the inward flux, which means the generalized force on the surface drives the flux from the surface into the bulk. If $\beta \mu_s <\mu_b$, $f_m<0$, then  $-\beta f_m$ is the outward flux.  
	\end{remark}
\begin{remark}
	In our previous work \cite{jing2022thermodynamically}, such a mobility is chosen
	\ben
	\bM_{4\times 4}=
	\left (
	\bea{lcrcr}
	M_c&0 & 0 & 0\\
	0 &M_s & \frac{\beta}{\alpha} & 0\\
	0&-\frac{\beta}{\alpha} & \frac{1}{\alpha} & 0
	\\
	0 &0 & 0& M_g
	\eea
	\right)=
	\left (
	\bea{lcrcr}
	M_c &0 & 0 & 0\\
	0 &M_s & 0 & 0\\
	0 &0 & \frac{1}{\alpha} & 0\\
	0 &0 & 0 & M_g
	\eea
	\right)+
	\left (
	\bea{lcrcr}
	0 &0 &0&0 &\\
	0 &0 & \frac{\beta}{\alpha} & 0\\
	0&-\frac{\beta}{\alpha} & 0 & 0\\
	0 &0 & 0 & 0
	\eea
	\right),\label{nonpurely}
	\een
	which is not the special case of \eqref{non-purely3}, since the govering system and boundary conditions does not satisfy the mass conservation law in the previous work, though the model can describe the reversible process. In fact, such mobility operaters are related to the reactive transport system. We will use another models to describe it in Section 4.4.
\end{remark}
\subsection{Model C for irreversible reactive transport system}
\noindent \indent In our previous work\cite{jing2022thermodynamically}, we developed a model for the reactive transport phase field model, where the reaction is describe an Allen-Cahn type equation and the transport process is described by a Cahn-Hilliard type equation. Such thermodynamically consistent model is given by
\bena
\parl{\phi}{t}=-[M_b^{(1)}-\nabla \cdot \bM_b^{(2)} \cdot \nabla]\mu_b,
\eena 
where $M_b^{(1)}$ is a special functional of $\phi$. In fact, the Allen-Cahn type mobility can also represent the source and sink terms. We restrict it as the reactive mobility due to the total mass conservation  for all species in this paper (or say $\phi_1+\phi_2=1,\psi(1)+\psi(2)=1$).  In the reactive transport system, the conservation law of mass of single species in \eqref{volumeevo1} will not be conserved, in meanwhile, the total mass of all species should be conserved. The consistent of model asks for that if the reaction vanishes by setting $M_b^{(1)}=0$, the model should also be correct.  In fact, the reaction will not affect the exchange of material directly, since the reaction occurs locally. However, the reaction will affect the exchange of energy and energy dissipation.

Based on the discussions in model A, the general surface mobility operator for irreversible reactive transport system is in \eqref{pure1}, which means model A is a special case of model C.
By setting $M_{11}^{(s)}=M_c, M_{14}^{(s)}=0,M_{14}^{(1)}=M_{24}^{(2)}=0$ and  $M_{44}=M_g$  for simplicity, the mobility operator on the surface should be

\ben
\bM_{4\times 4}=
\left (
\bea{lcrcr}
M_{c}  &M_{12}^{(1)}-\nabla_s\cdot (M_{12}^{(2)}\nabla_s) &0 &0\\
M_{12}^{(1)}-\nabla_s\cdot (M_{12}^{(2)}\nabla_s) & \frac{\beta^2}{\alpha}+M_{22}^{(1)}-\nabla_s\cdot (M_{22}^{(2)}\nabla_s) & -\frac{\beta}{\alpha}  & 0\\
0 & -\frac{\beta}{\alpha} & \frac{1}{\alpha} & 0
\\
0& 0 & 0 & M_{g}
\eea
\right),
\een
where $M_{12}^{(1)}$ and $M_{22}^{(1)}$ are two mobility for reactions. The governing system of equations together with the dynamic boundary conditions is summarized as follows.
\bena\label{g-model3}
\phi_t=-M_b^{(1)}\mu_b+\nabla \cdot \bM_b^{(2)} \cdot \nabla \mu_b,  \quad \bx \in \Omega, \\
\phi_t=-M_c \mu_c-M_{12}^{(1)} \mu_s+\nabla_s\cdot (M_{12}^{(2)}\nabla_s \mu_s), \quad \bs \in \partial \Omega, \\
\psi_t=-M_{12}^{(1)} \mu_c+\nabla_s\cdot (M_{12}^{(2)}\nabla_s \mu_c)-[M_{22}^{(1)}-\nabla_s\cdot (M_{22}^{(2)}\nabla_s)]\mu_s-\beta f_m, \quad \bs \in \partial \Omega, \\
\alpha \bn \cdot \bM_b^{(2)} \cdot \nabla \mu_b=-\mu_b+\beta \mu_s, \quad
\nabla_\bn\phi_t=-M_g \mu_g,  \quad \bs \in \partial \Omega. 
\eena

The corresponding  energy dissipation rate is given by
\ben
\bea{l}
\frac{dE}{dt}=- \int_{\Omega} \mu_b  M_b^{(1)} \mu_b+\nabla \mu_b  \cdot \bM_b^{(2)} \cdot \nabla \mu_b d\bx-
\int_{\partial \Omega} \bU \cdot \bM_{4\times 4} \cdot \bU d\bs\leq 0,\label{energy-diss1}
\eea
\een
where
$
\bU=
\left (
\bea{lcrcr}
\phi_t,&
\psi_t,&
f_m,&
\nabla_{\bn}\phi_t
\eea
\right)^T
$, $\bM_{4\times 4}$ is semi-positive definite.

\subsection{Model D for irreversible-reversible reactive transport system} 
\noindent \indent For a reactive transport system including both irreversible and reversible processes, the mass conservation law is only valid for all species of the system. The surface mobility operator in \eqref{non-purely1} is the general case for irreversible-reversible reactive transport system, which means model B is a special case of model D. When setting $M_{12}^{(a)}=M_{14}^{(s)}=M_{14}^{(a)}=M_{24}^{(a)}=M_{24}^{(2)}=M_{34}^{(s)}=0, M_{11}^{(s)}=M_c,M_{44}^{(s)}=M_g$ for simplicity, then the mobility of the surface is given by
\bena
\bM_{4\times 4} =\left (
\bea{lcrcr}
M_{c}  &M_{12}^{(s)} &2M_{13}^{(s)} &0\\
M_{12}^{(s)}& M_{22}^{(s)} & \frac{\beta}{\alpha}+2M_{23}^{(s)}  & 0\\
0 & -\frac{\beta}{\alpha} & \frac{1}{\alpha} & 0
\\
0& 0 & 0 & M_{g}
\eea
\right), \label{nonpurely3}
\eena
where $M_{12}^{(s)}=M_{12}^{(1)}-\nabla_s \cdot (M_{12}^{(2)}\nabla_s),M_{22}^{(s)}=M_{22}^{(1)}-\nabla_s \cdot (M_{22}^{(2)}\nabla_s),M_{23}^{(s)}=M_{23}^{(1)}-\nabla_s \cdot (M_{23}^{(2)}\nabla_s)$.
The governing system of equations together with the dynamic boundary conditions in this model is summarized as follows,
\bena\label{g-model4}
\phi_t=-M_b^{(1)}\mu_b+\nabla \cdot \bM_b^{(2)} \cdot \nabla \mu_b,  \quad \bx \in \Omega, \\
\phi_t=-M_c \mu_c-M_{12}^{(s)} \mu_s-2M_{13}^{(s)}\mu_b,\quad \bs \in \partial \Omega, \\
\psi_t=-M_{12}^{(s)} \mu_c-M_{22}^{(s)}\mu_s-(\frac{\beta}{\alpha}+2M_{23}^{(s)} )\mu_b, \quad \bs \in \partial \Omega, \\
\alpha \bn \cdot \bM_b^{(2)} \cdot \nabla \mu_b=-\mu_b+\beta \mu_s, \quad
\nabla_\bn\phi_t=-M_g \mu_g,  \quad \bs \in \partial \Omega. 
\eena
The corresponding  energy dissipation rate is given by
\ben
\bea{l}
\frac{dE}{dt}=- \int_{\Omega} \mu_b M_b^{(1)}\mu_b+\nabla \mu_b  \cdot \bM_b^{(2)} \cdot \nabla \mu_b d\bx-
\int_{\partial \Omega} \bU \cdot \bM_{4\times 4}^{(s)} \cdot \bU d\bs\leq 0,\label{energy-diss2}
\eea
\een
where
$
\bU=
\left (
\bea{lcrcr}
\phi_t,&
\psi_t,&
f_m,&
\nabla_{\bn}\phi_t
\eea
\right)^T
$, $\bM_{4\times 4}^{(s)}$ is semi-positive definite,  $\bM_{4\times 4}^{(a)}$ does not contribute to the energy dissipation rate but affect the exchange of material between the bulk and surface.
\begin{remark}
The mobility in our previous work \cite{jing2022thermodynamically}
	\ben
	\bM_{4\times 4}=
	\left (
	\bea{lcrcr}
	M_c&0 & 0 & 0\\
	0 &M_s & \frac{\beta}{\alpha} & 0\\
	0&-\frac{\beta}{\alpha} & \frac{1}{\alpha} & 0
	\\
	0 &0 & 0& M_g
	\eea
	\right)=
	\left (
	\bea{lcrcr}
	M_c &0 & 0 & 0\\
	0 &M_s & 0 & 0\\
	0 &0 & \frac{1}{\alpha} & 0\\
	0 &0 & 0 & M_g
	\eea
	\right)+
	\left (
	\bea{lcrcr}
	0 &0 &0&0 &\\
	0 &0 & \frac{\beta}{\alpha} & 0\\
	0&-\frac{\beta}{\alpha} & 0 & 0\\
	0 &0 & 0 & 0
	\eea
	\right),
	\een
	which is a special case of \eqref{nonpurely3} by setting $M_{22}=M_s, M_{12}=M_{23}=0, $, the mass of the first species is not conserved in this model.
\end{remark}

\section{Numerical simulation}
\noindent \indent  Since the reaction occurs locally and does not affect the exchange of material between the bulk and surface directly, we focus on models A and B. In this section, we will use structure-preserving algorithms, energy quadratic and finite difference methods on stagger grid  \cite{zhao2018general,gong2018second} to perform simulations and compare the difference of Model A in \eqref{g-model1} and Model B in \eqref{g-model2} numerically. The bulk and surface energy densities are given by
\bena
e_b=\frac{\gamma_1}{2}|\nabla \phi|^2+\phi \ln \phi+(1-\phi) \ln (1-\phi)+\chi_b \phi (1-\phi),\\
e_s=\frac{\gamma_1}{2}|\nabla_s \psi|^2+\psi \ln \psi+(1-\psi) \ln (1-\psi)+\chi_s \psi (1-\psi), \\
 e_c=0.
\eena  We list a model as following and setting a dynamic boundary condition on one of the surface $\Gamma$ and homogeneous Neumann boundary conditions in the other boundaries $\partial \Omega /\Gamma$.
\bena
\phi_t=\nabla \cdot \bM_b^{(2)} \cdot \nabla \mu_b,  \quad \bx \in \Omega, \\
\phi_t=-M_c \mu_c+\nabla_s \cdot (M_{12}^{(2)}\nabla_s \mu_s)\textcolor{red}{-2M_{13}^{(s)}\mu_b},\quad \bs \in \Gamma, \\
\psi_t=\nabla_s \cdot (M_{12}^{(2)}\nabla_s \mu_c)+\nabla_s \cdot (M_{22}^{(2)}\nabla_s\mu_s)+\textcolor{red}{2\nabla_s \cdot (M_{23}^{(2)}\nabla_s \mu_b)}-\beta f_m, \quad \bs \in \Gamma, \\
\alpha \bn \cdot \bM_b^{(2)} \cdot \nabla \mu_b=-\mu_b+\beta \mu_s, \quad
\nabla_\bn\phi_t=-M_g \mu_g,  \quad \bs \in \Gamma,\\
\bn \cdot \nabla \phi=\bn \cdot \nabla \mu_b=0, \quad \bs\in \partial \Omega /\Gamma,
\eena
The red terms in the equations describe the reversible transport process. If $M_{23}^{(2)}=M_{13}^{(s)}=0$, the model reduces to Model A for pure irreversible transport process. However, once $M_{23}^{(2)}\neq 0$ or $M_{13}^{(s)}\neq 0$, model A and model B are different models with different dynamics and energy dissipation rates.
We set the computation domain $\Omega=(-1,1)^2$ with $C^6$ smoothness at the corner, $M_b^{(2)}=5\times 10^{-6},M_{c}=1\times 10^{-6},M_{12}^{(s)}=1\times 10^{-6},M_{13}^{(s)}=0,M_{22}^{(s)}=1\times 10^{-5},M_{23}^{(s)}=1\times 10^{-5},\chi_b=4,\chi_s=5,\gamma_1=b^2/3,\gamma_2=b^2/6,b=0.02$, the time step size $\Delta t=1\times 10^{-4}, $ the spatial step size $\Delta h=1/128$  and the initial condition is given by 
\bena
\phi(\bx)=0.3+0.01*\zeta,\quad \phi(\bs)=\lim_{\bx \rightarrow \partial \Omega} \phi(\bx), \quad \psi(\bs)=\phi(\bs), \quad \zeta \sim U [-1,1].
\eena

\begin{figure*}
			\subfigure[]{
			\begin{minipage}[b]{0.24\linewidth}
				\includegraphics[width=1\linewidth]{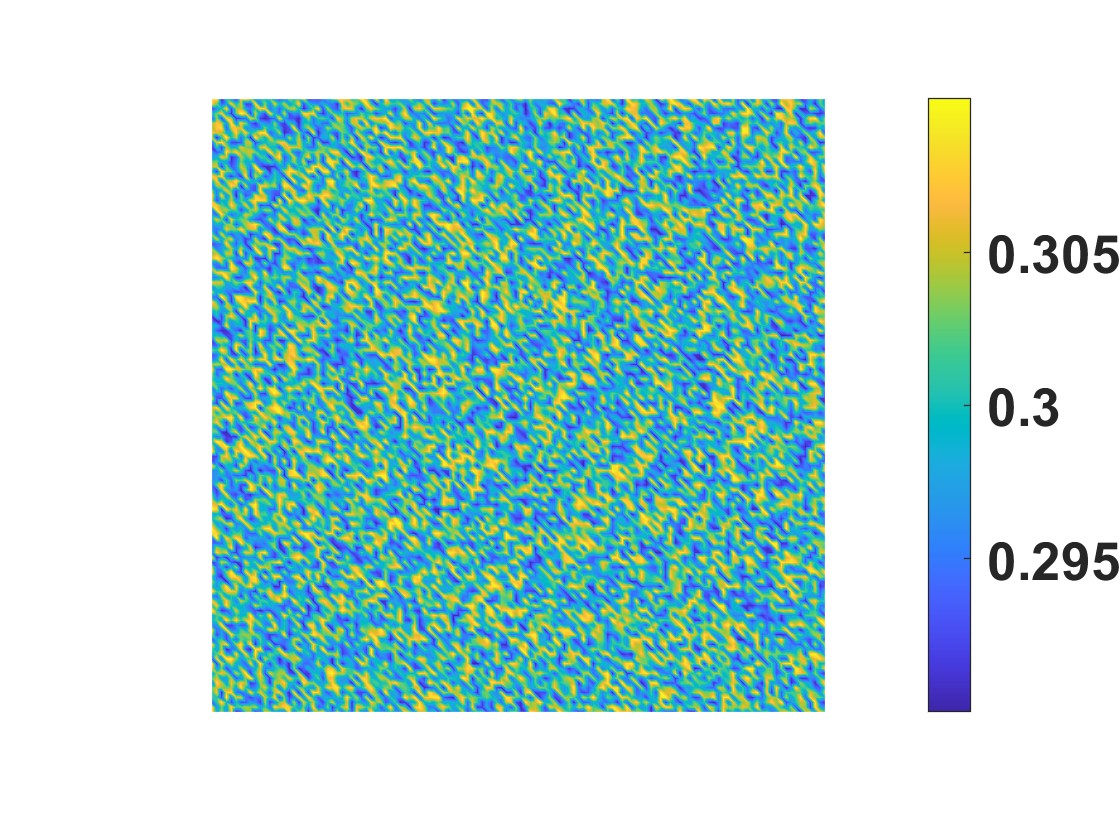}
		\end{minipage}}
		\subfigure[]{
			\begin{minipage}[b]{0.24\linewidth}
				\includegraphics[width=1\linewidth]{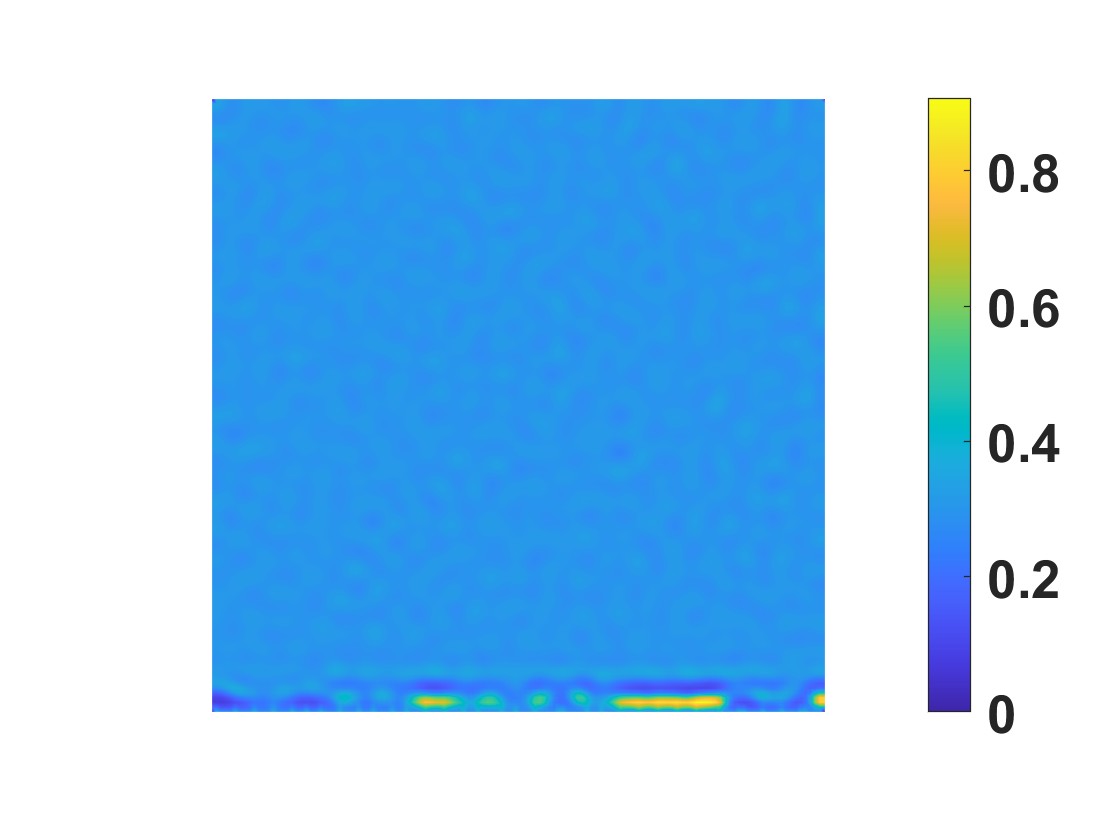}
		\end{minipage}}
		\subfigure[]{
			\begin{minipage}[b]{0.24\linewidth}
				\includegraphics[width=1\linewidth]{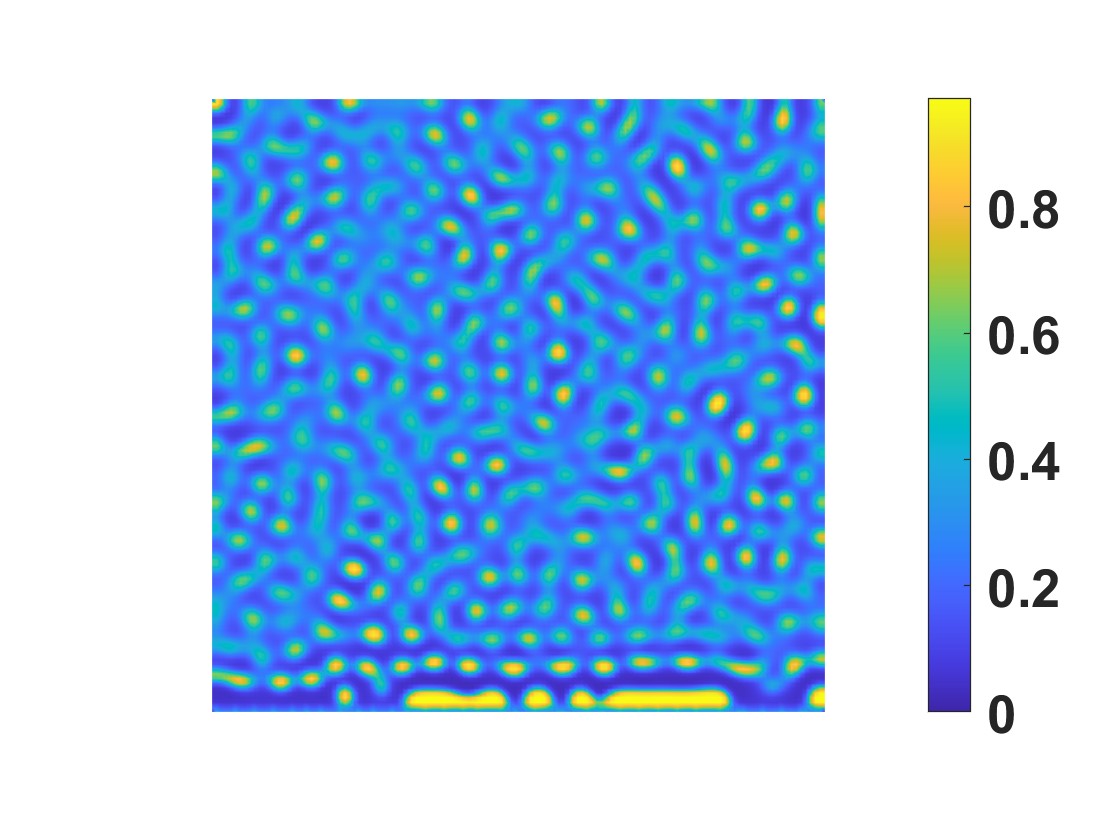}
		\end{minipage}}
		\subfigure[]{
			\begin{minipage}[b]{0.24\linewidth}
				\includegraphics[width=1\linewidth]{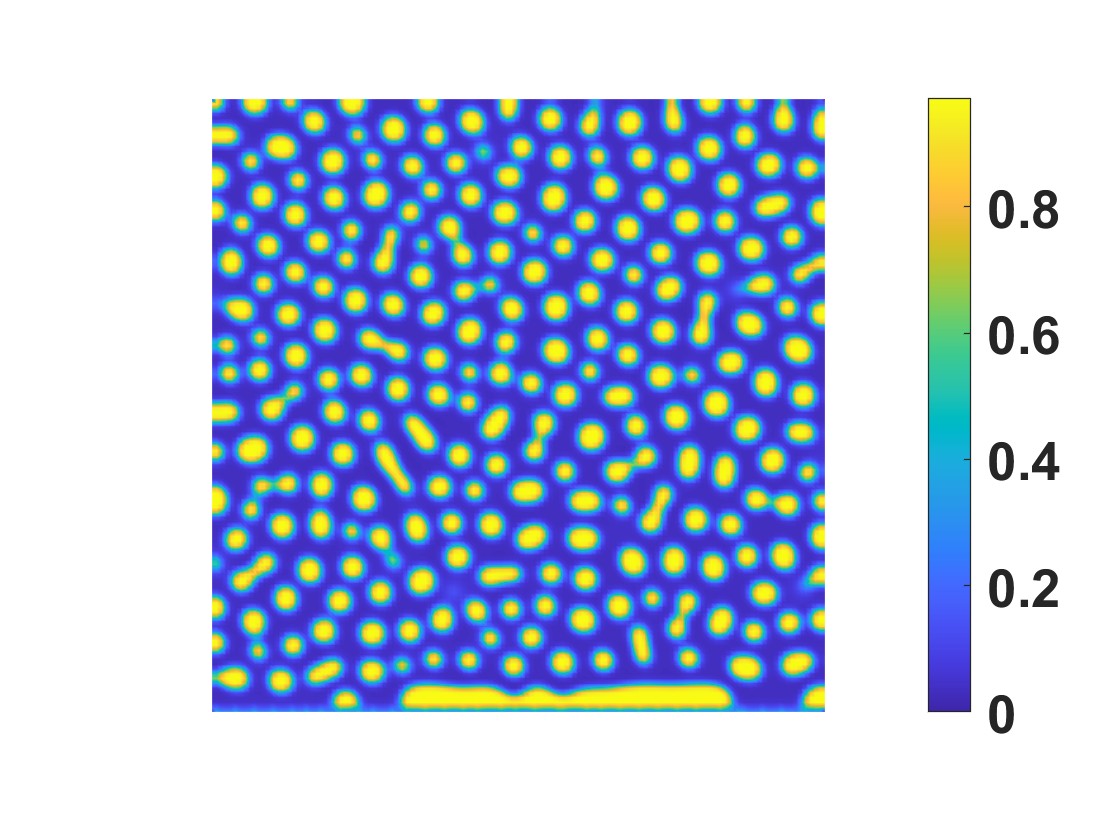}
		\end{minipage}}\\
		\subfigure[]{
		\begin{minipage}[b]{0.24\linewidth}
			\includegraphics[width=1\linewidth]{T=0.jpg}
	\end{minipage}}
		\subfigure[]{
			\begin{minipage}[b]{0.24\linewidth}
				\includegraphics[width=1\linewidth]{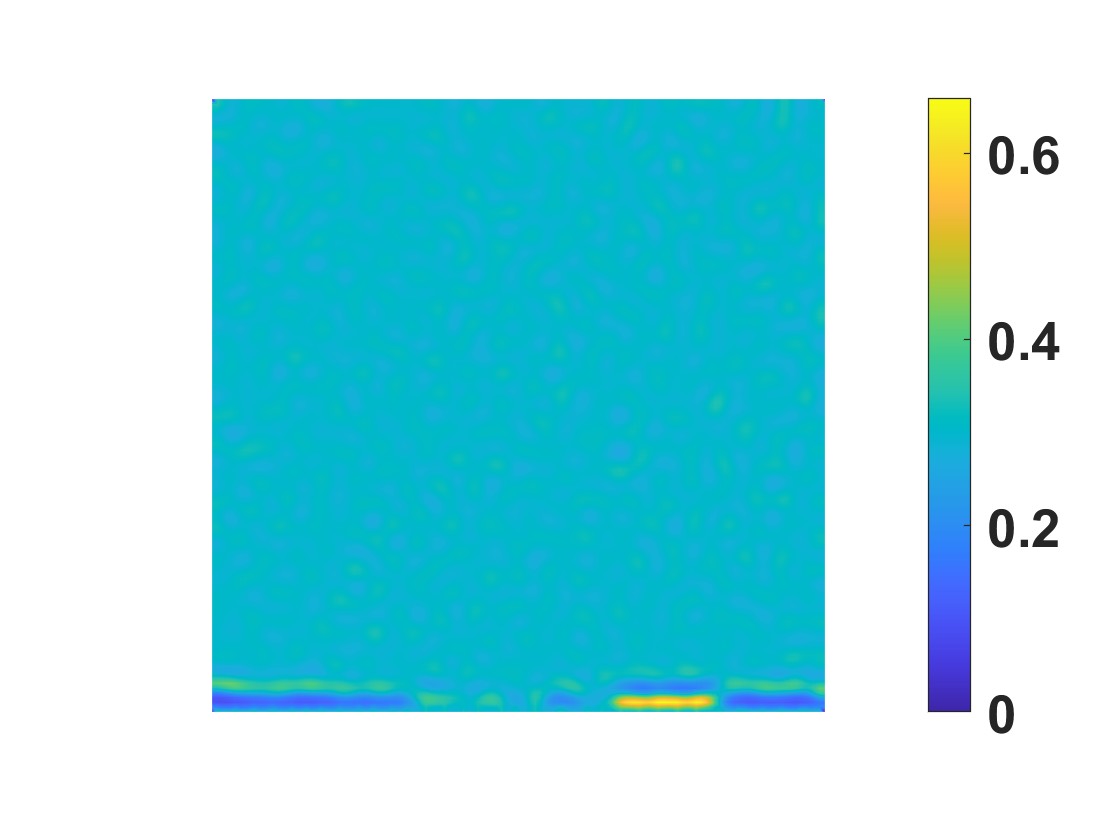}
	\end{minipage}}
\subfigure[]{
\begin{minipage}[b]{0.24\linewidth}
\includegraphics[width=1\linewidth]{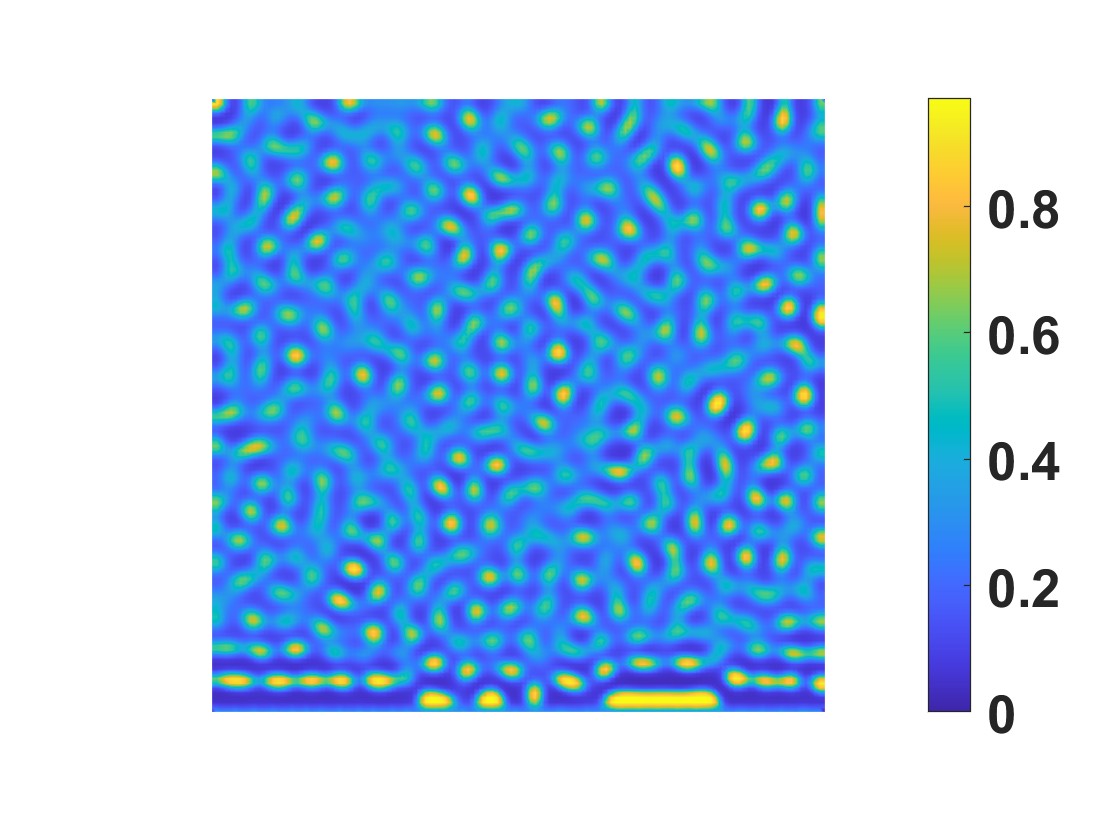}
\end{minipage}}
\subfigure[]{
\begin{minipage}[b]{0.24\linewidth}
\includegraphics[width=1\linewidth]{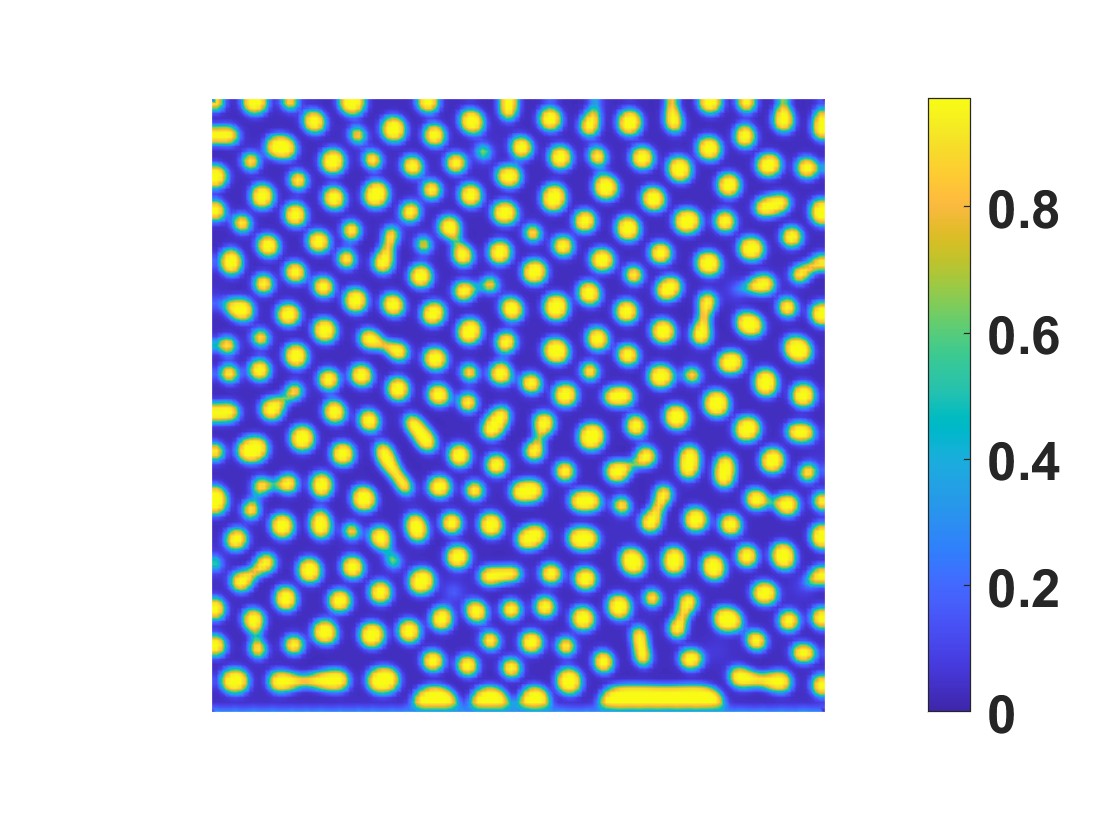}
\end{minipage}}
\caption{Time evolution of the spinodal decomposition of the binary fluid system with $M_{23}=0$ (a-d) and $M_{23}=1\times 10^{-5}$ (e-h), respectively.   Snapshots of numerical solutions $\phi$ are taken at $t=0,20,50,100$ for the two cases, respectively.   Dynamic boundary effects on the bulk solution near $\Gamma$ are observed in both cases and $M_{23}^{(s)} $ can affect the phase separation significantly.}\label{figure1}
\end{figure*}

\begin{figure*}
	\subfigure[]{
		\begin{minipage}[b]{0.33\linewidth}
			\includegraphics[width=1\linewidth]{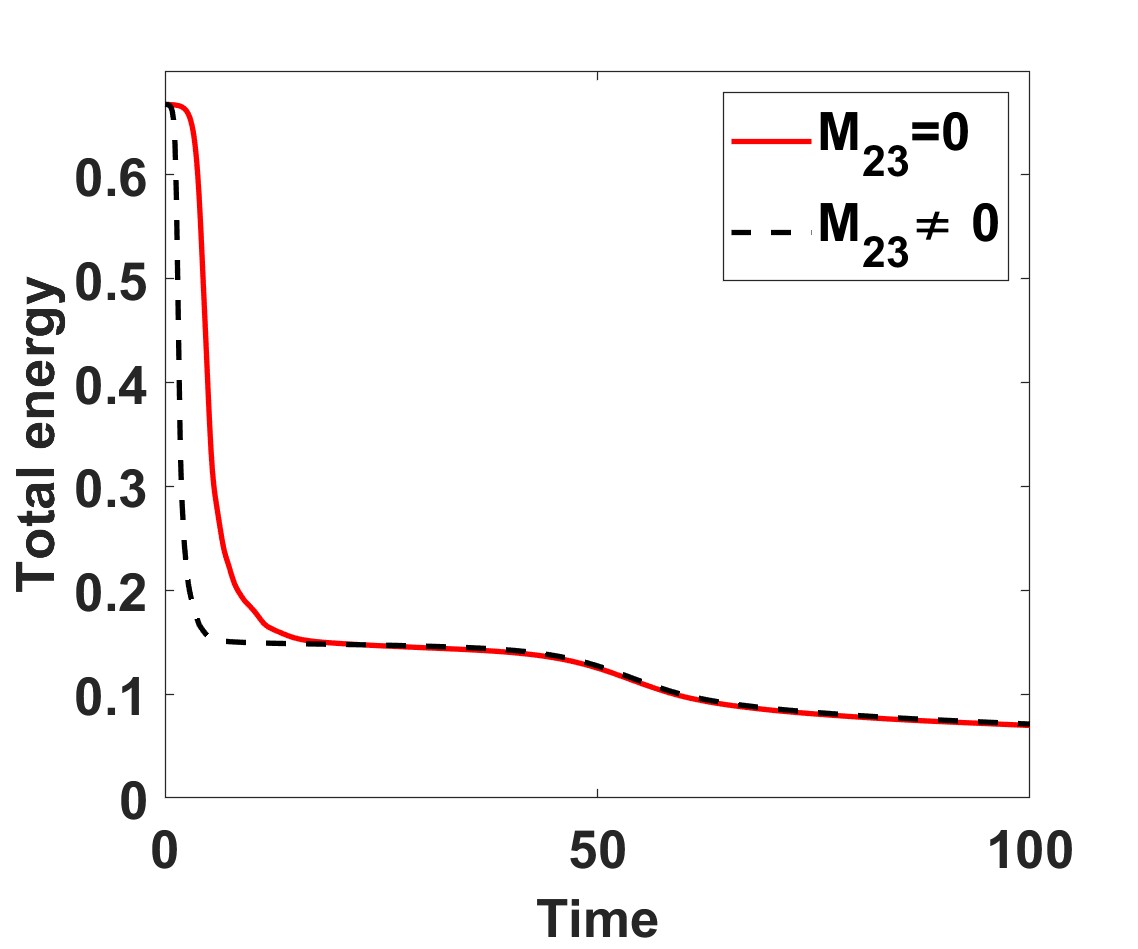}
	\end{minipage}}
	\subfigure[]{
		\begin{minipage}[b]{0.33\linewidth}
			\includegraphics[width=1\linewidth]{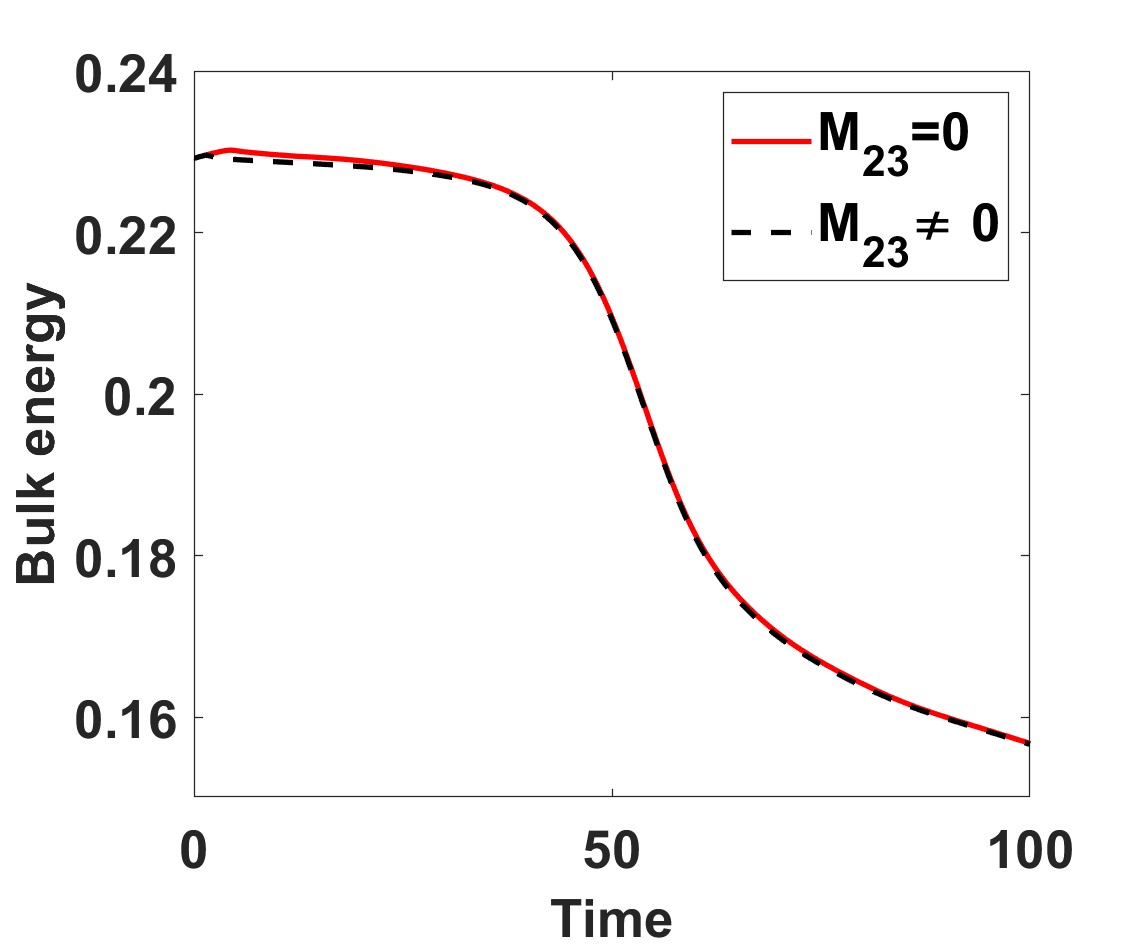}
	\end{minipage}}
	\subfigure[]{
		\begin{minipage}[b]{0.33\linewidth}
			\includegraphics[width=1\linewidth]{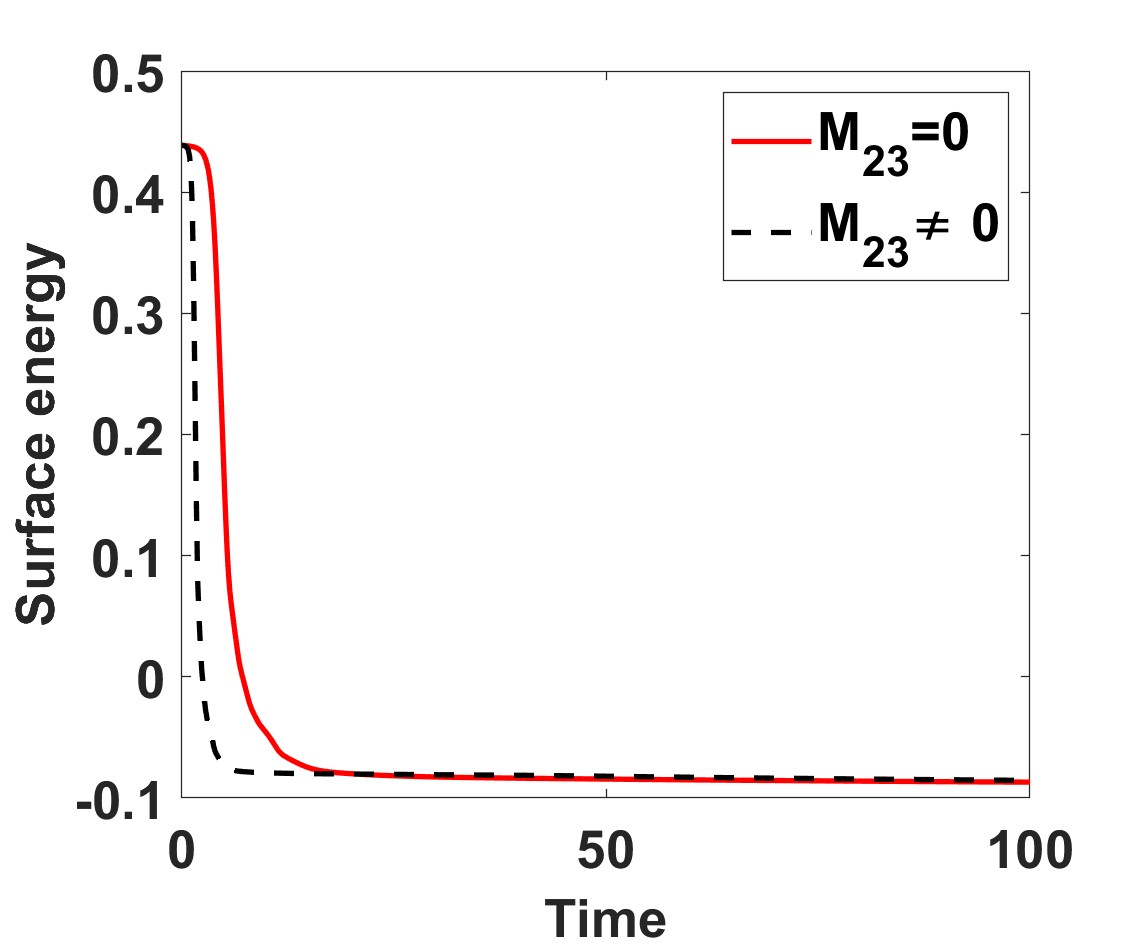}
	\end{minipage}}
	\caption{Time evolutions of the free energy in total, free energy in bulk and free energy on surface, the free energy in total dissipates with time, however, the free energy in bulk is not monotonic decreasing with time. The numerical results are coincident with our theory.  }\label{figure2}
\end{figure*}

\begin{figure*}
	\subfigure[]{
		\begin{minipage}[b]{0.33\linewidth}
			\includegraphics[width=1\linewidth]{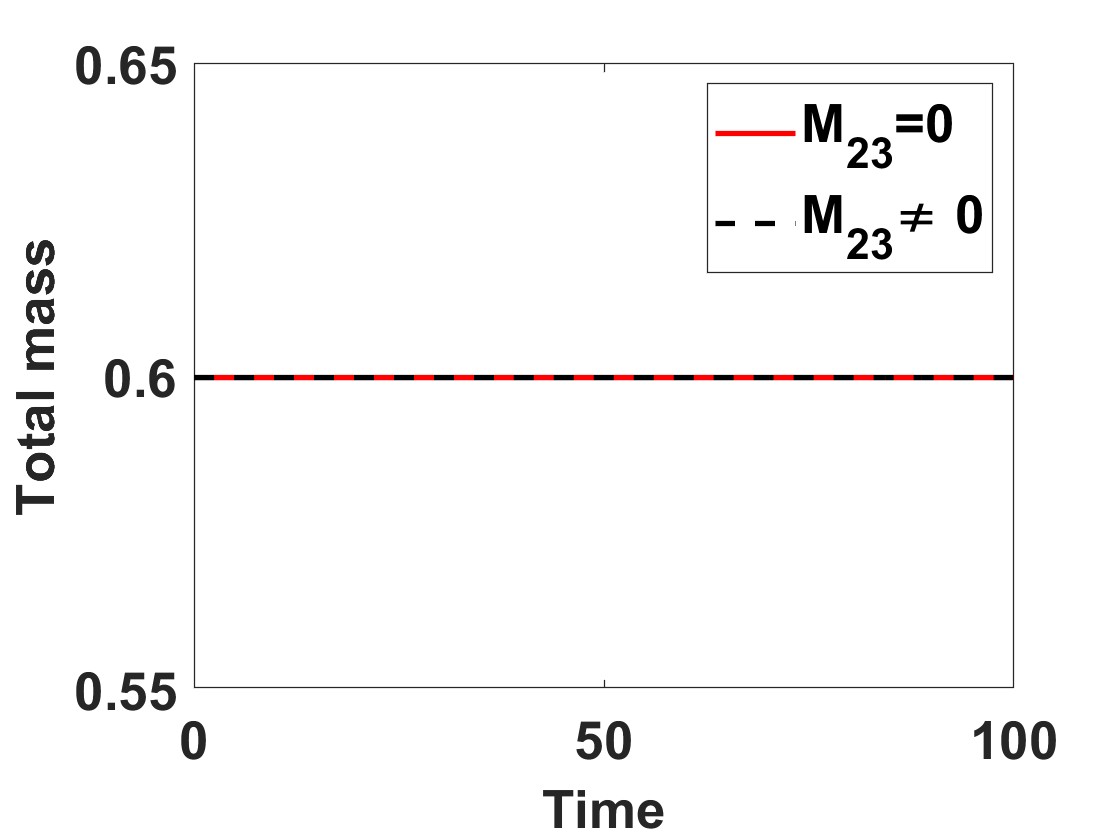}
	\end{minipage}}
	\subfigure[]{
		\begin{minipage}[b]{0.33\linewidth}
			\includegraphics[width=1\linewidth]{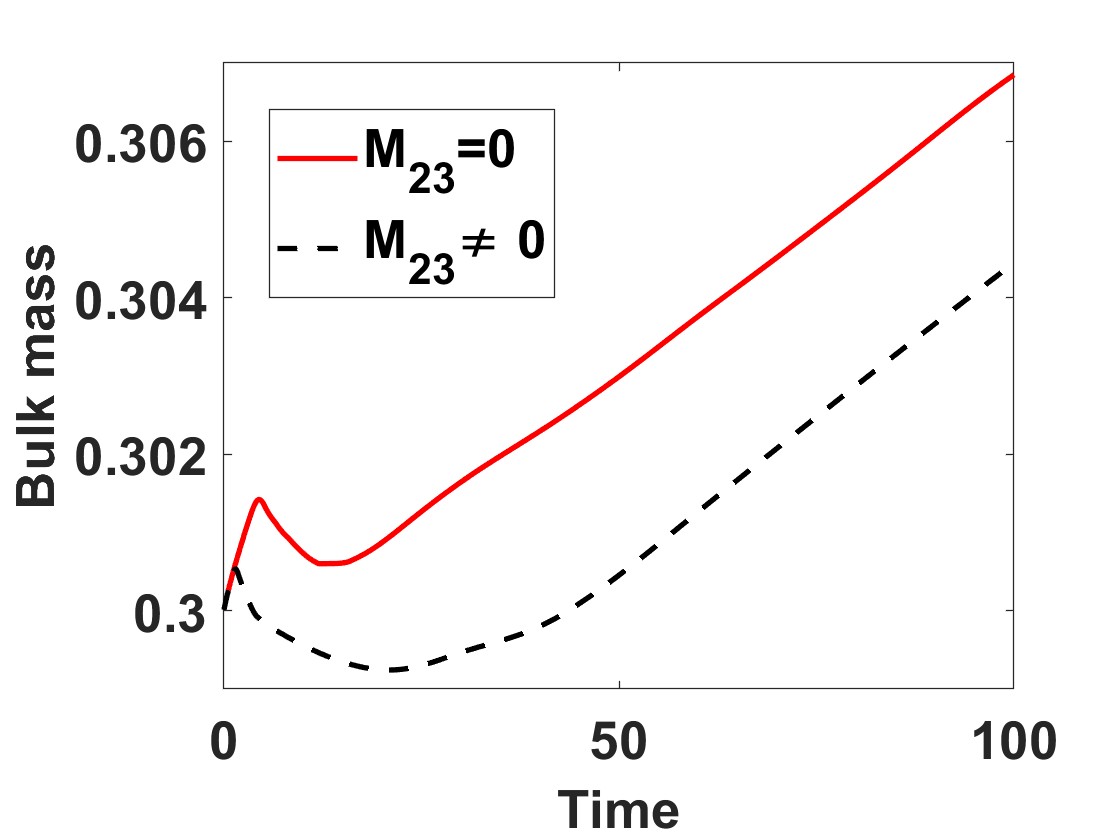}
	\end{minipage}}
	\subfigure[]{
		\begin{minipage}[b]{0.33\linewidth}
			\includegraphics[width=1\linewidth]{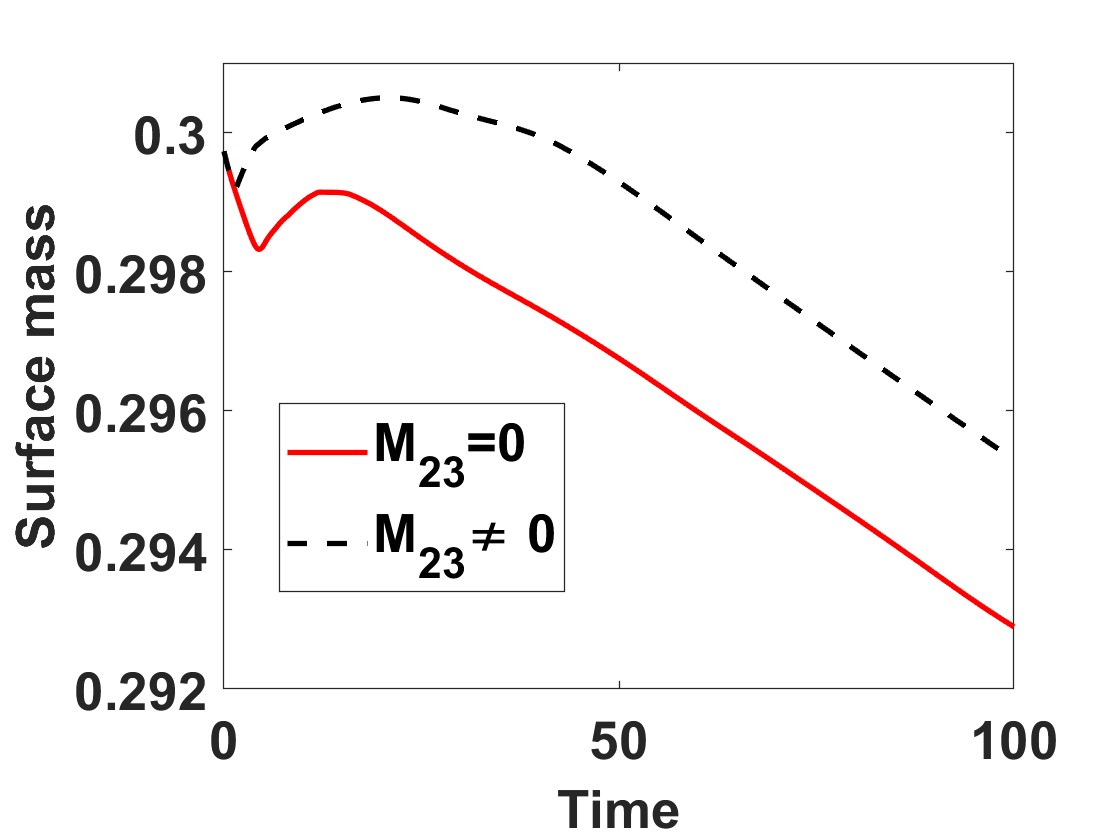}
	\end{minipage}}
	\caption{Time evolutions of the mass in total, bulk mass and surface mass, the mass in total is conserved, in meanwhile, the mass in bulk and on the surface are not monotonic with time. The numerical results are coincident with our theory, this is attribute to the structure-preserving numerical algorithms are used.  }\label{figure3}
\end{figure*}

We then use the structure preserving numerical methods to simulate the exchange of mass and energy between the bulk and surface.
We now present numerical examples to illustrate the impact of boundary dynamics on bulk evolution. The first simulation (Figure~\ref{figure1}a–d) depicts spinodal decomposition with $M_{23}^{(s)}=0$ in Figure~\ref{figure1} (a--d) at $t=0,20,50,100$. The second simulation (Figure~\ref{figure1}e–h) repeats this under identical conditions but with $M_{23}^{(s)}=1 \times 10^{-5}$. The nonzero  parameter $M_{23}^{(s)}$ value modifies the pattern morphology and accelerates the energy dissipation in Figure~\ref{figure2}. Comparisons of bulk and surface free energies (Figure~\ref{figure2}-b) reveal that the bulk energy does not decrease monotonically when 
$M_{23}^{(s)}=0$, in accordance with theoretical expectations. While total mass remains conserved in both cases (Figure~\ref{figure3}), bulk and surface masses vary individually due to interfacial exchange. These comparisons confirm that dynamic boundary conditions noticeably influence bulk dynamics via non-negligible flux across the surface, affecting both mass transport and energy dissipation.

\begin{figure*}
	\subfigure[]{
		\begin{minipage}[b]{0.33\linewidth}
			\includegraphics[width=1\linewidth]{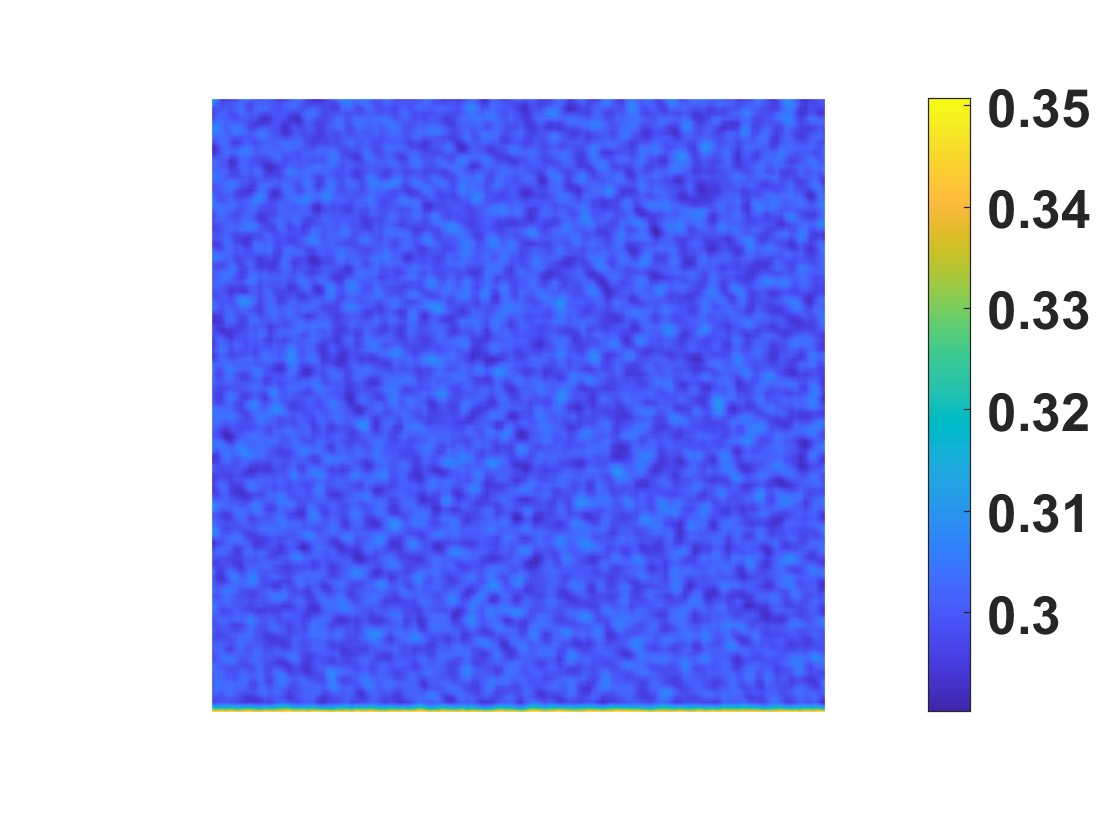}
	\end{minipage}}
	\subfigure[]{
		\begin{minipage}[b]{0.33\linewidth}
			\includegraphics[width=1\linewidth]{T=50.jpg}
	\end{minipage}}
	\subfigure[]{
		\begin{minipage}[b]{0.33\linewidth}
			\includegraphics[width=1\linewidth]{T=100.jpg}
	\end{minipage}}\\
	\subfigure[]{
		\begin{minipage}[b]{0.33\linewidth}
			\includegraphics[width=1\linewidth]{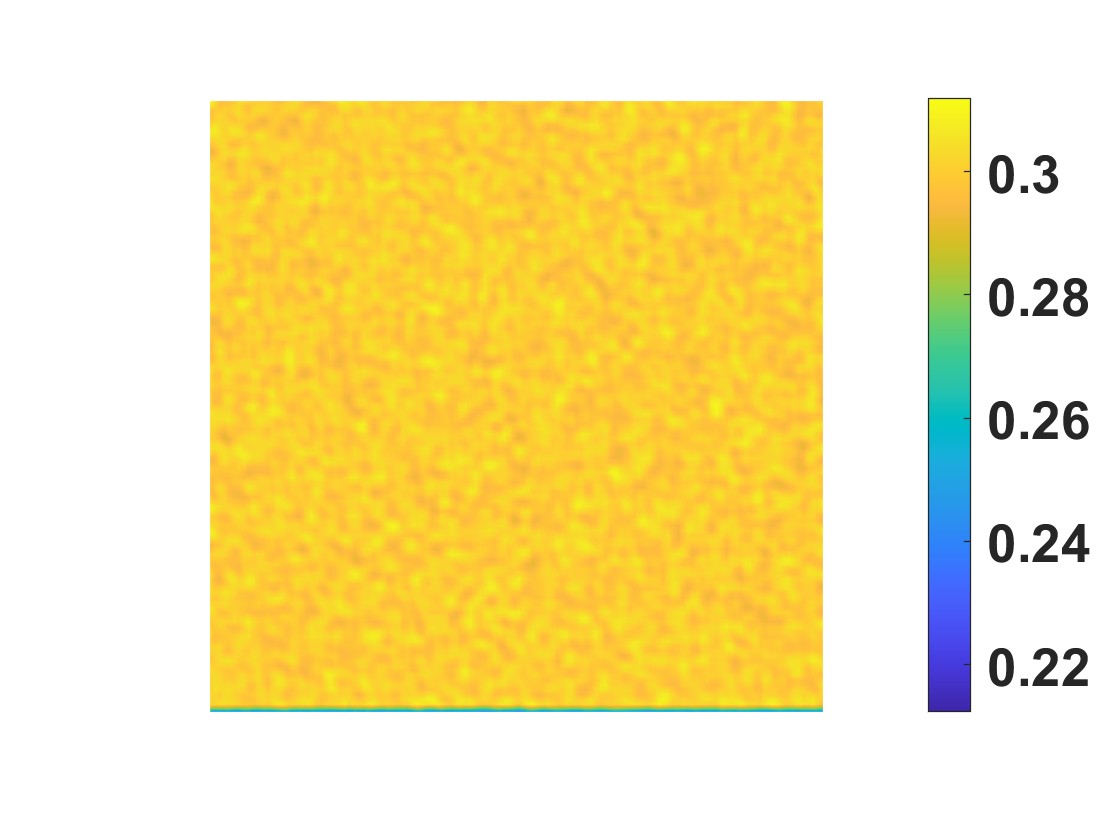}
	\end{minipage}}
	\subfigure[]{
		\begin{minipage}[b]{0.33\linewidth}
			\includegraphics[width=1\linewidth]{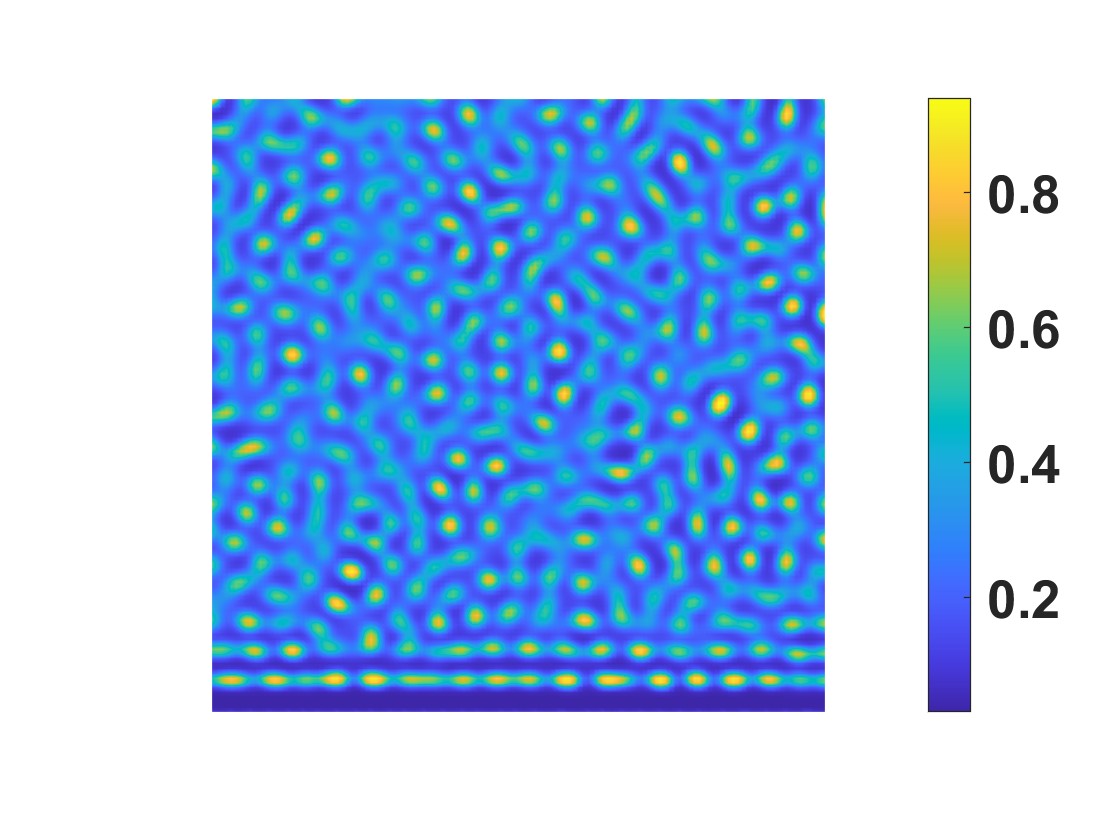}
	\end{minipage}}
	\subfigure[]{
		\begin{minipage}[b]{0.33\linewidth}
			\includegraphics[width=1\linewidth]{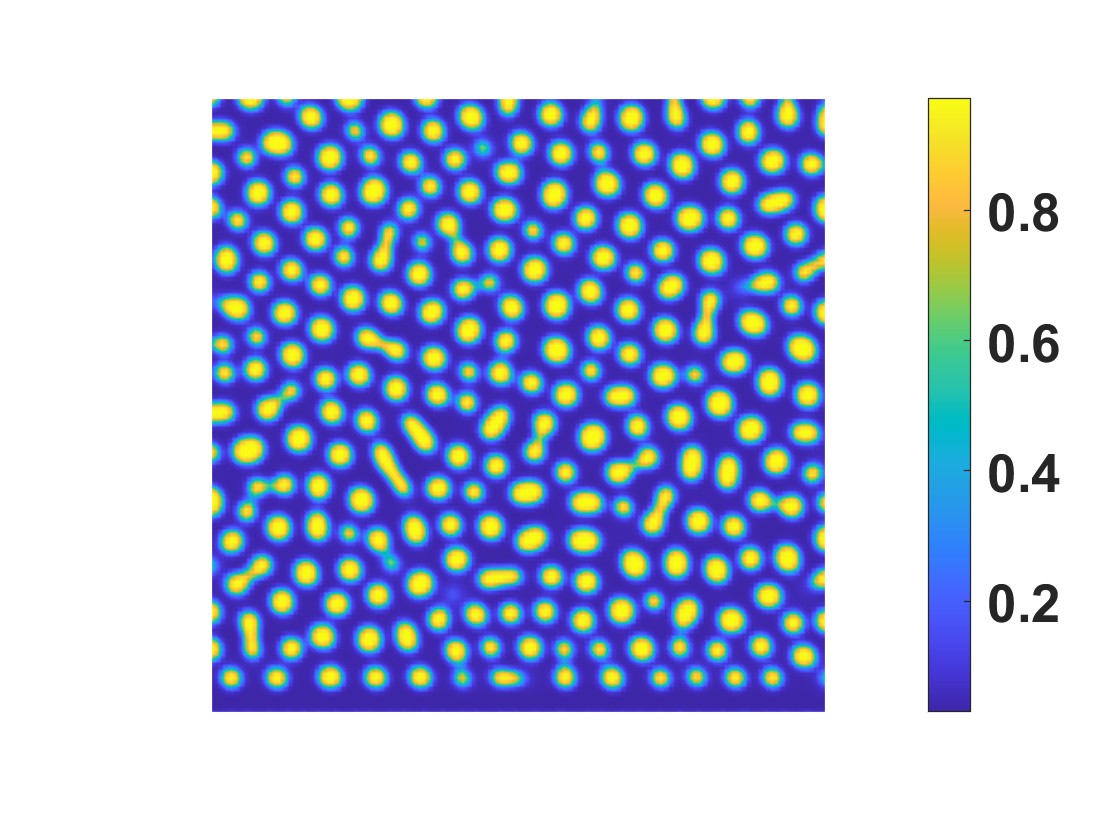}
	\end{minipage}}\\
		\subfigure[]{
		\begin{minipage}[b]{0.33\linewidth}
			\includegraphics[width=1\linewidth]{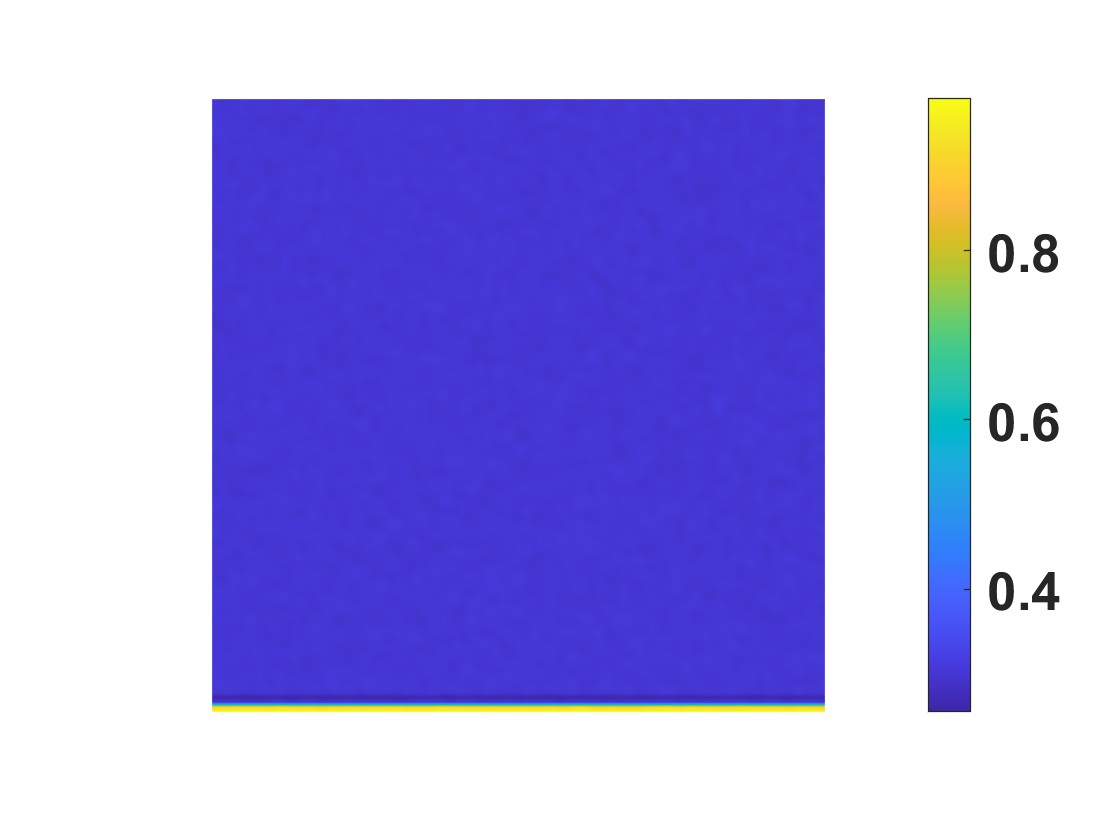}
	\end{minipage}}
	\subfigure[]{
		\begin{minipage}[b]{0.33\linewidth}
			\includegraphics[width=1\linewidth]{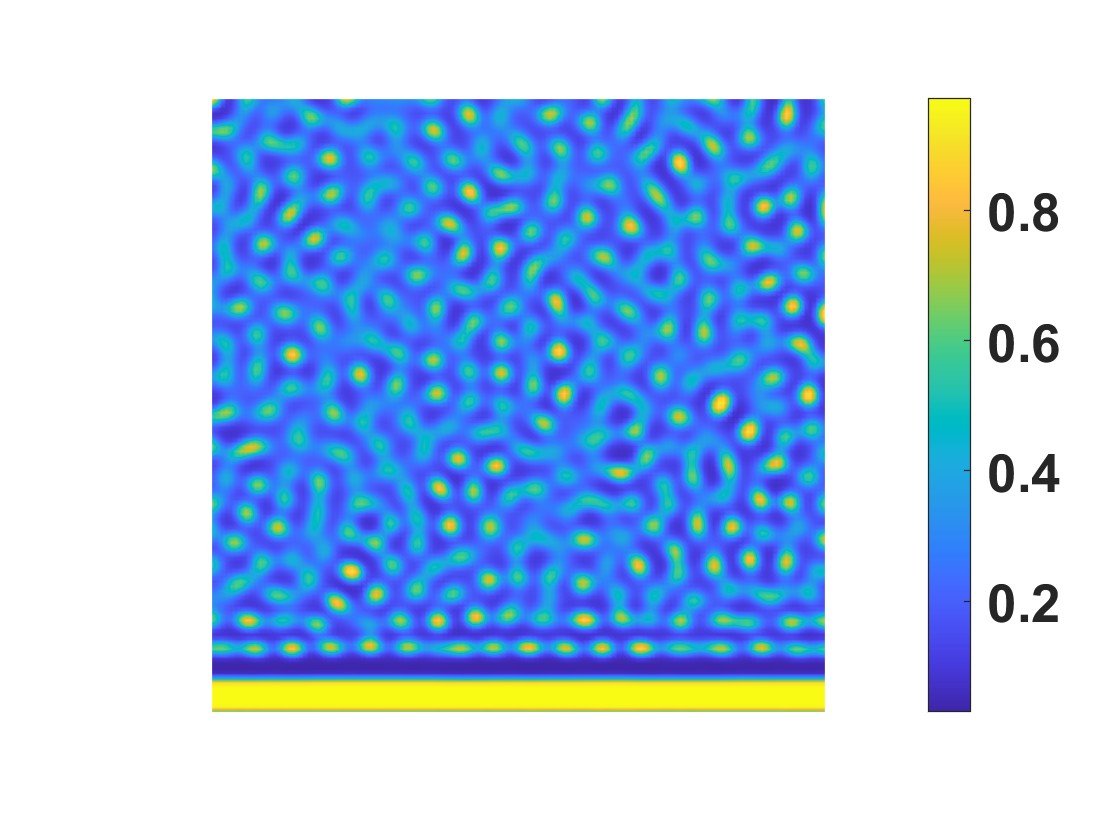}
	\end{minipage}}
	\subfigure[]{
		\begin{minipage}[b]{0.33\linewidth}
			\includegraphics[width=1\linewidth]{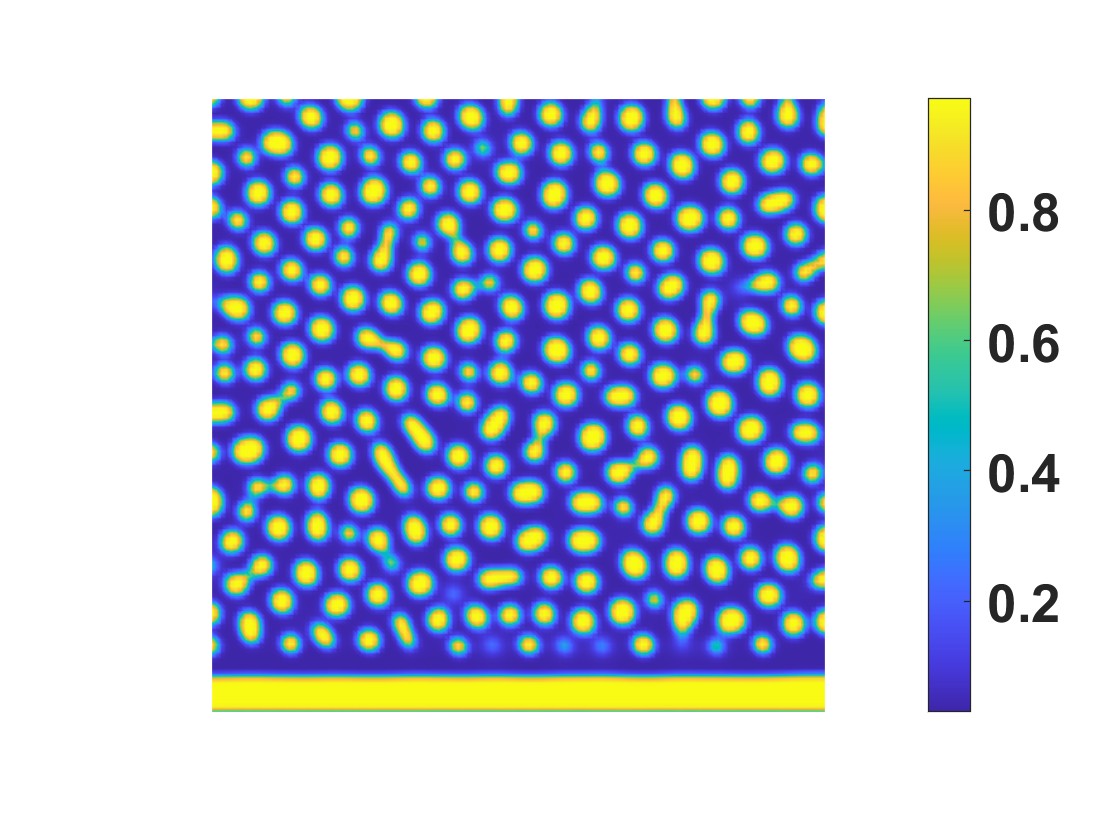}
	\end{minipage}}\\
	\caption{Time evolutions of the spinodal decomposition of the binary fluid system with $\beta =1$ in (a-c), $\beta =1\times 10^{-3}$ in (d-f) and $\beta =10$ in (g-i) at $t=7,50,100$, respectively. The effects of characteristic length on the patterns are observed in these three cases.}\label{figure4}
\end{figure*}

\begin{figure*}
	\subfigure[]{
		\begin{minipage}[b]{0.33\linewidth}
			\includegraphics[width=1\linewidth]{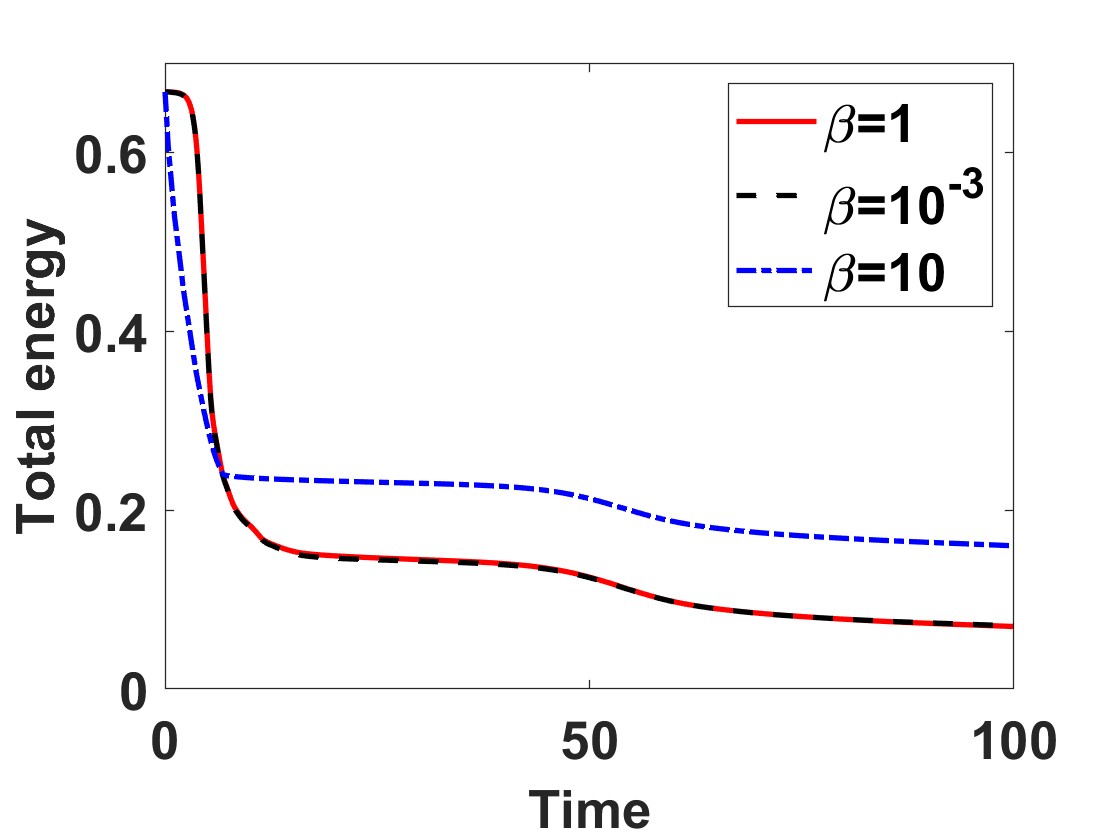}
	\end{minipage}}
	\subfigure[]{
		\begin{minipage}[b]{0.33\linewidth}
			\includegraphics[width=1\linewidth]{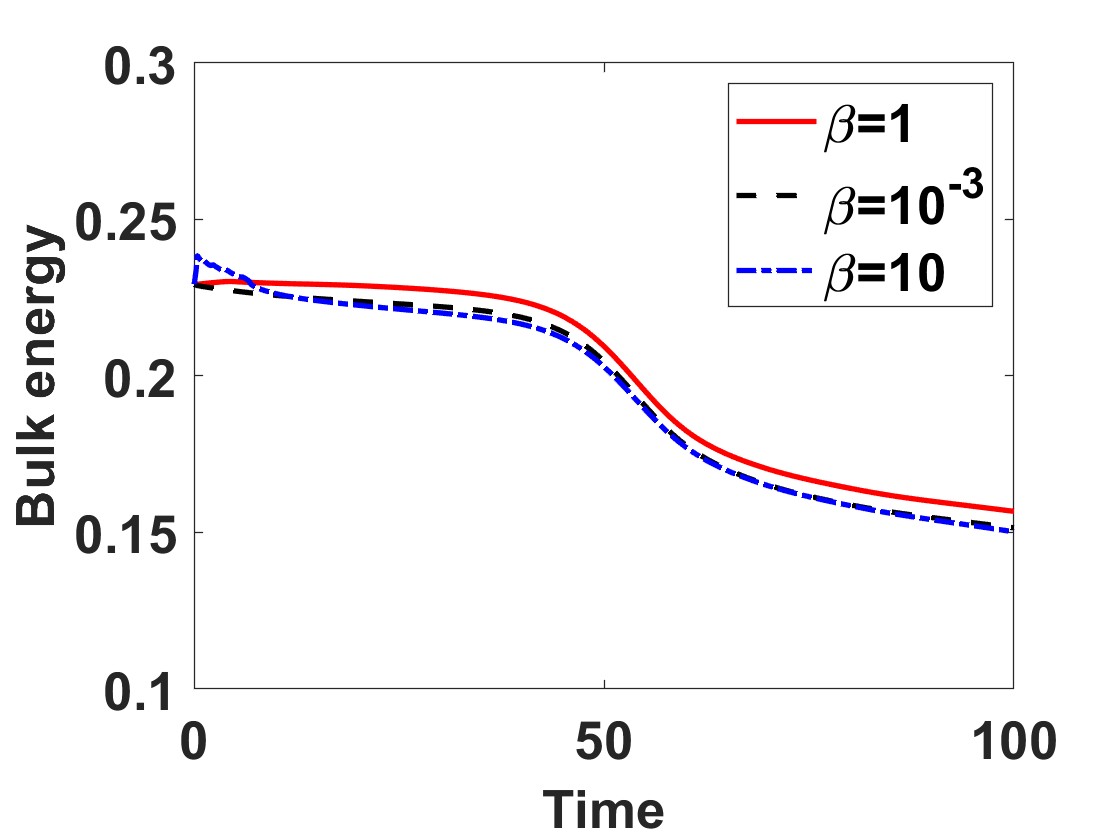}
	\end{minipage}}
	\subfigure[]{
		\begin{minipage}[b]{0.33\linewidth}
			\includegraphics[width=1\linewidth]{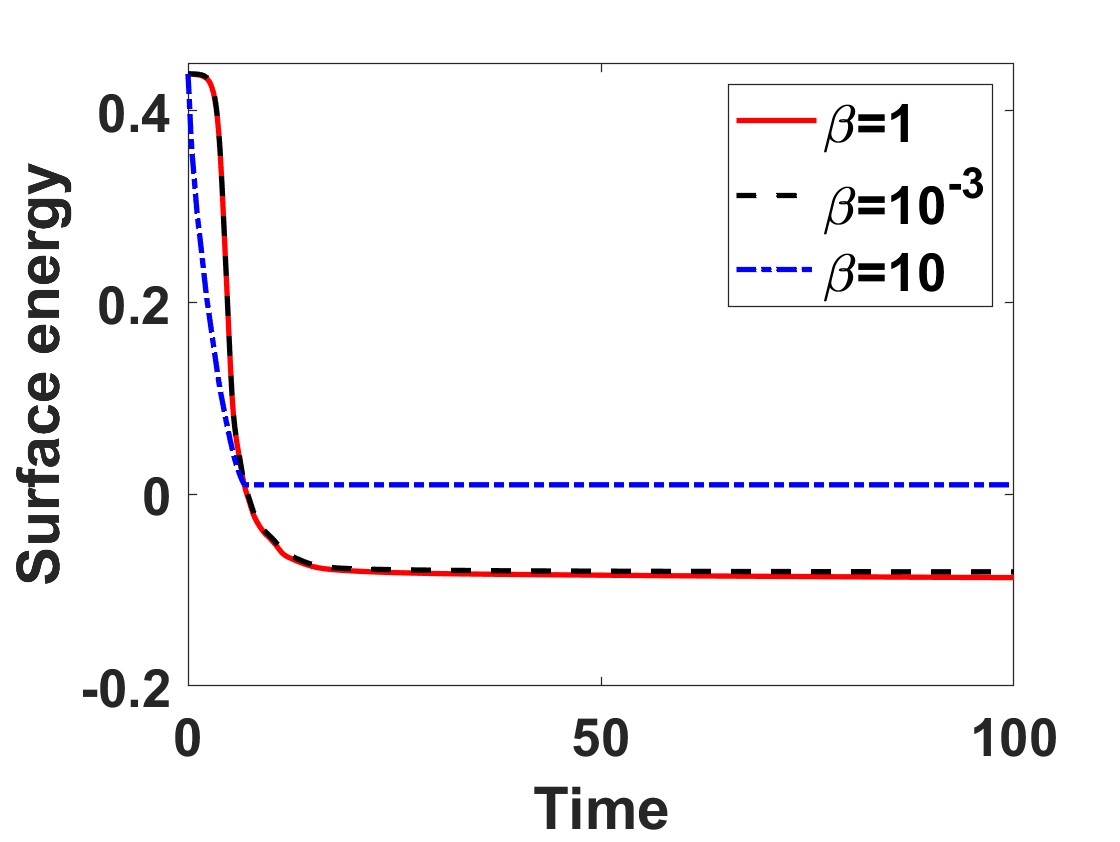}
	\end{minipage}}
	\caption{Time evolutions of the free energy in total, free energy in bulk and free energy on surface with $\beta=1, \beta=1\times 10^{-3}, 10$. The total energy is dissipated in three cases, the bulk enengy and surface energy are not always monotonic with time (for example, the bulk energy with $\beta=10$ increases at the first stage and then decreases with time). The numerical results are coincident with our theory.  }\label{figure5}
\end{figure*}

\begin{figure*}
	\subfigure[]{
		\begin{minipage}[b]{0.33\linewidth}
			\includegraphics[width=1\linewidth]{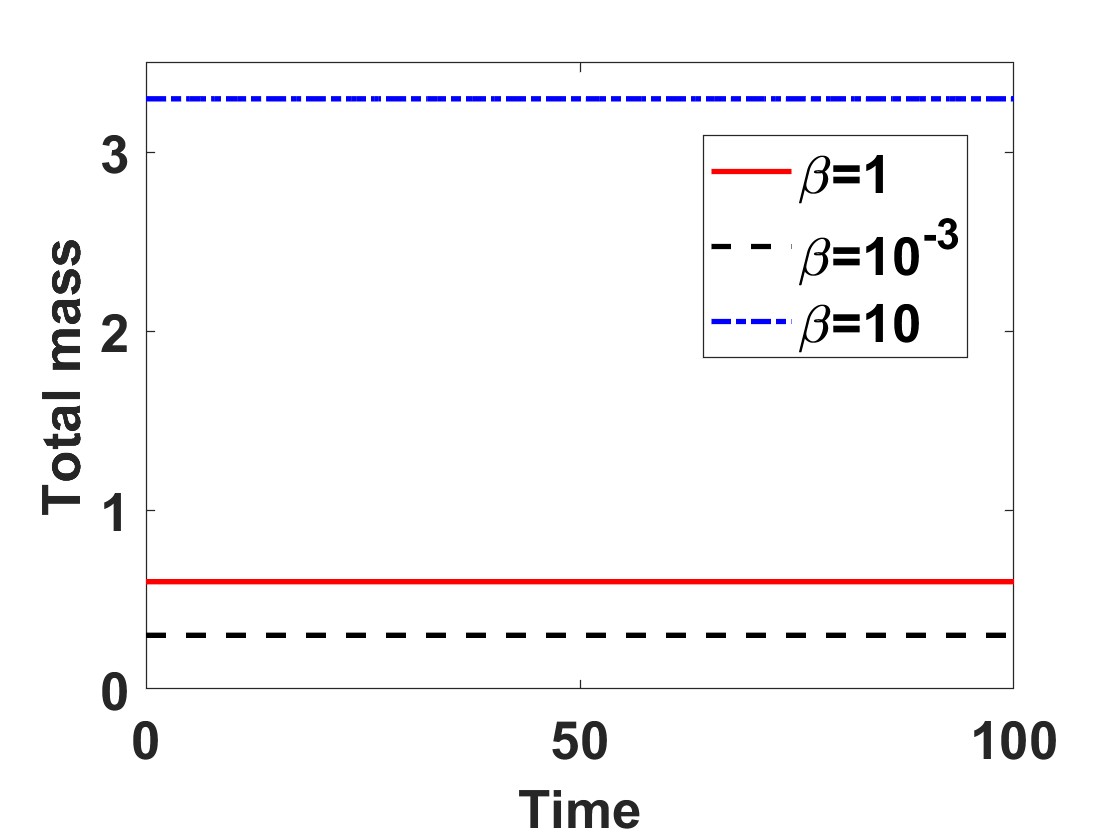}
	\end{minipage}}
	\subfigure[]{
		\begin{minipage}[b]{0.33\linewidth}
			\includegraphics[width=1\linewidth]{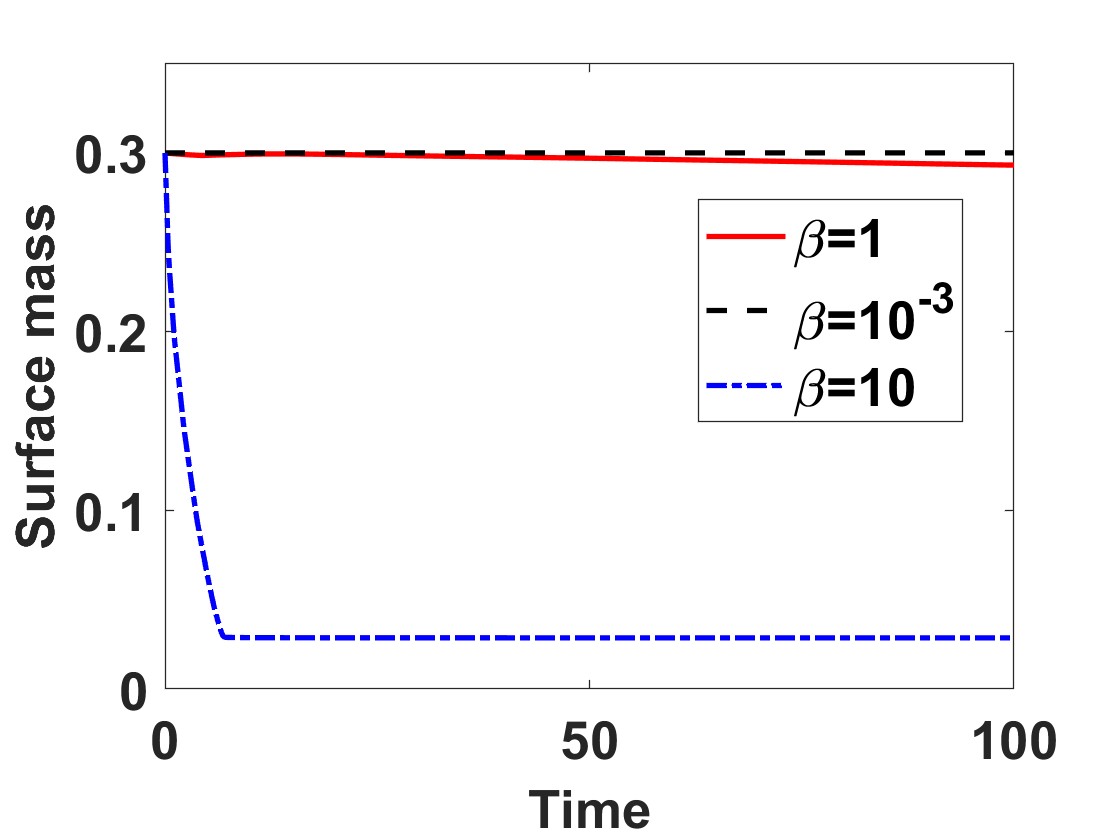}
	\end{minipage}}
	\subfigure[]{
	\begin{minipage}[b]{0.33\linewidth}
		\includegraphics[width=1\linewidth]{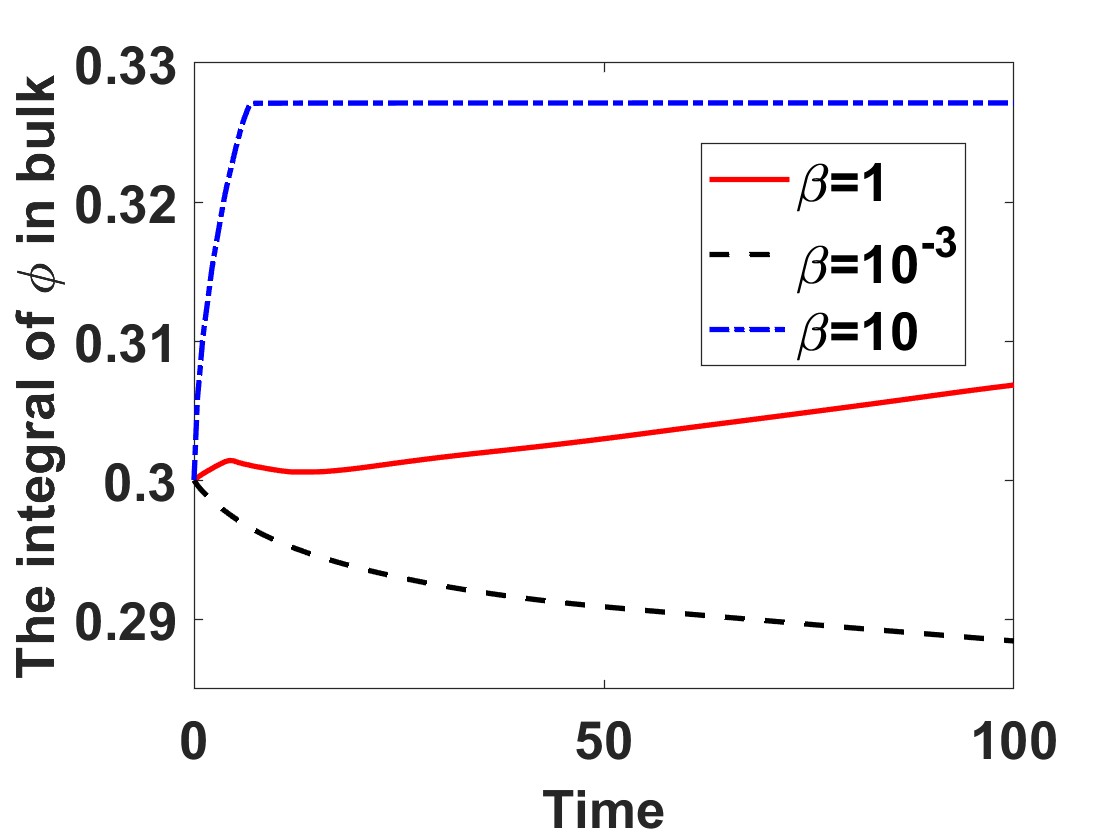}
\end{minipage}}
	\subfigure[]{
		\begin{minipage}[b]{0.33\linewidth}
			\includegraphics[width=1\linewidth]{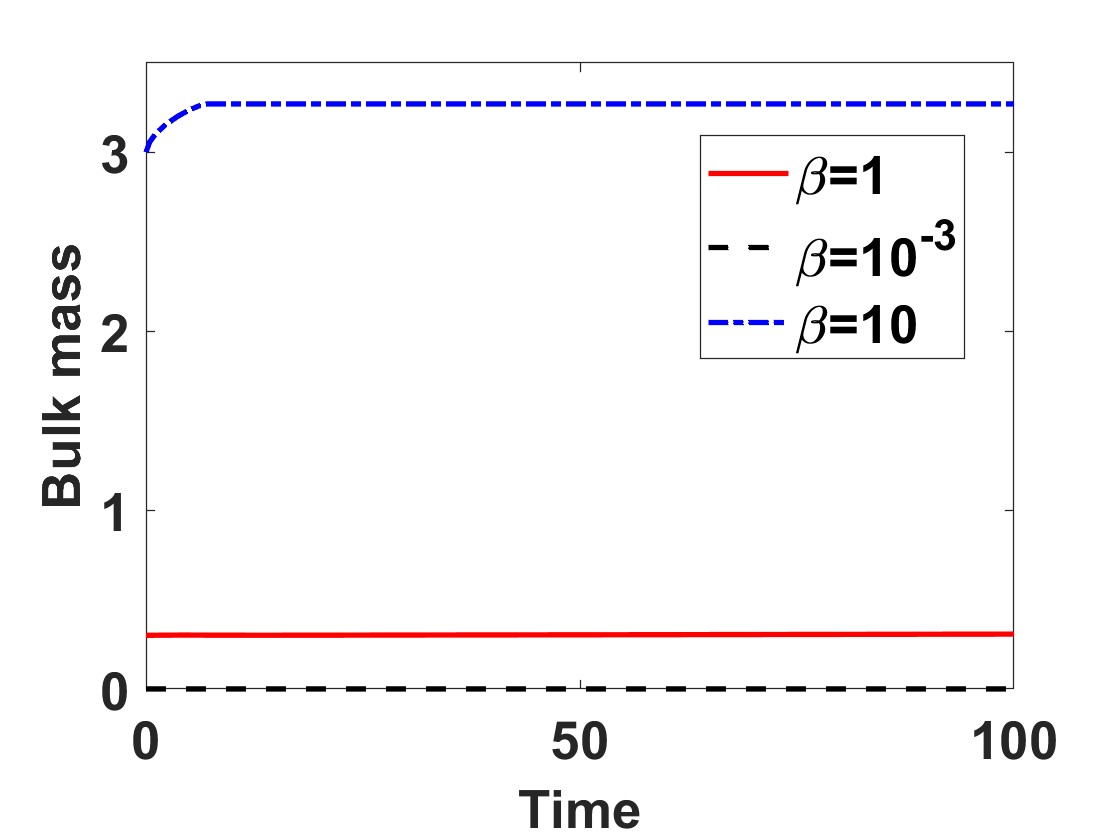}
	\end{minipage}}
		\subfigure[]{
		\begin{minipage}[b]{0.33\linewidth}
			\includegraphics[width=1\linewidth]{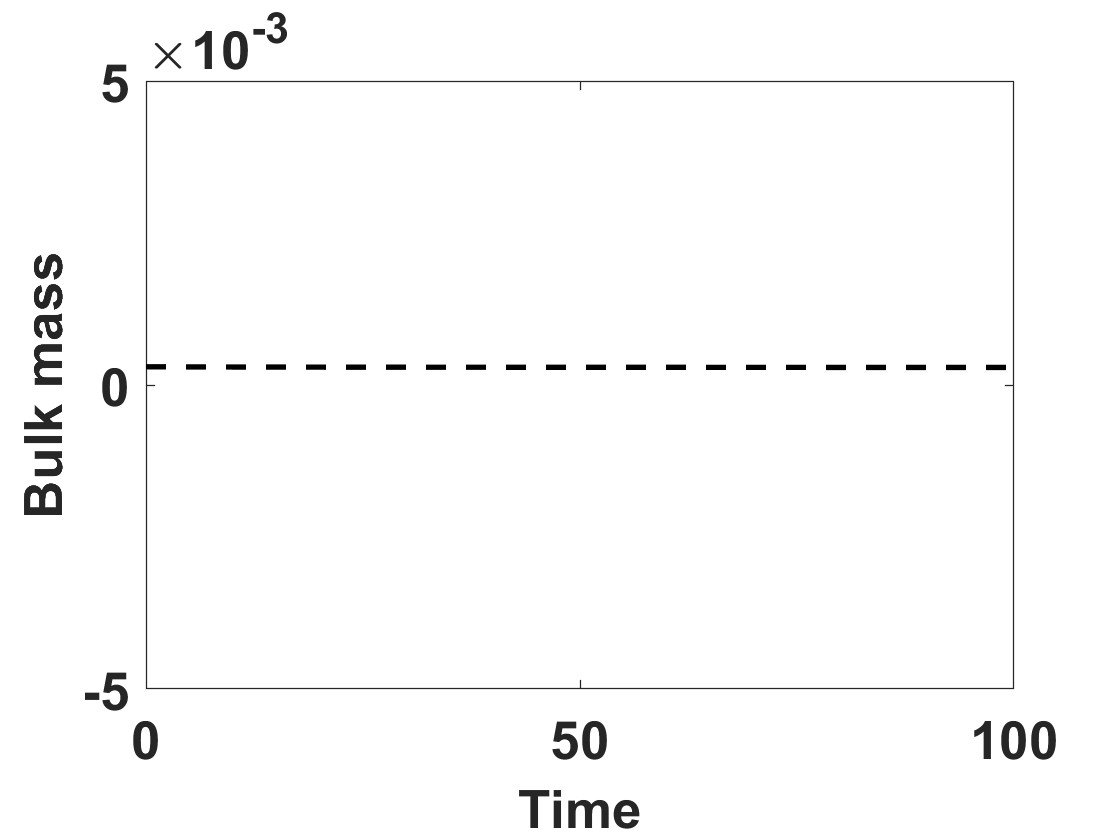}
	\end{minipage}}
	\subfigure[]{
	\begin{minipage}[b]{0.33\linewidth}
		\includegraphics[width=1\linewidth]{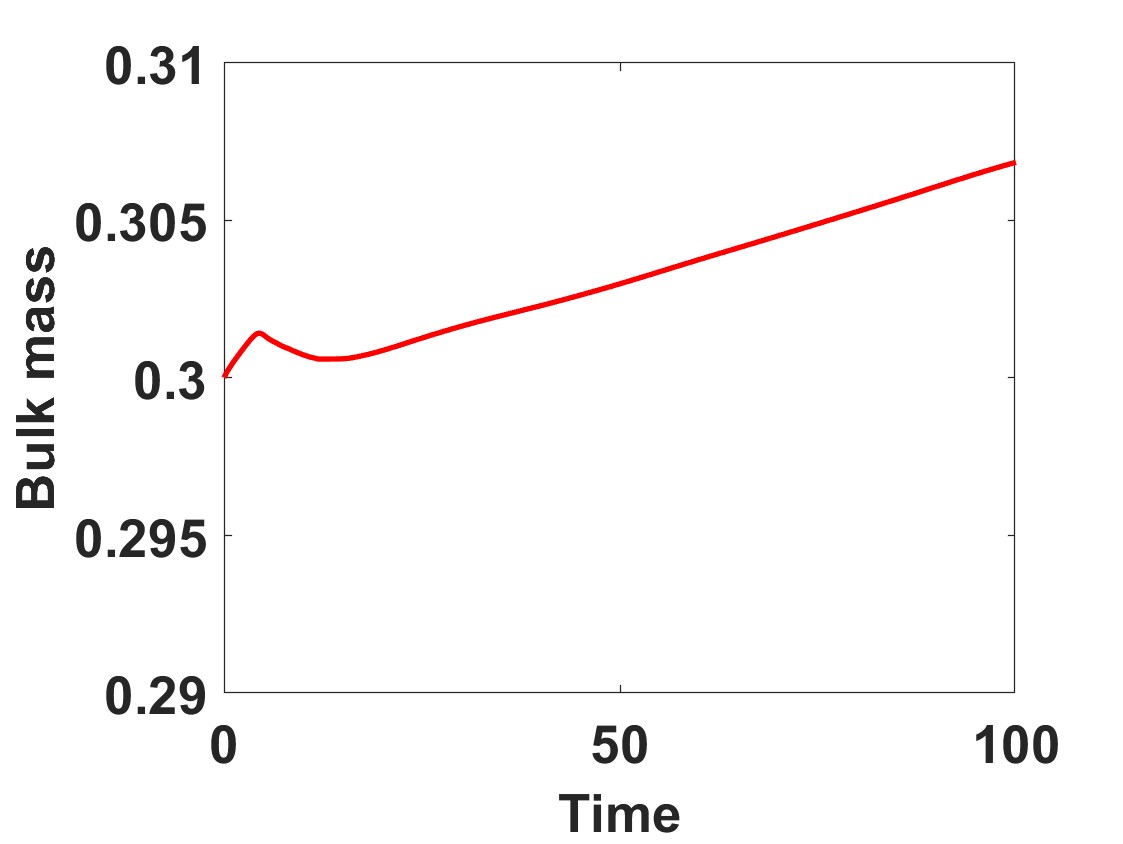}
\end{minipage}}
	\caption{Time evolutions of the total mass, surface mass and bulk mass with $\beta=1, 1\times 10^{-3}, 10$. The total masses in three cases are conserved in (a), the surface masses in (b) and the bulk integrals of $\phi$ in (c)  are not always conserved, due to the effects of exchange of material. The bulk mass in (d) are also not conserved with time, (e-f) are the time evolutions of bulk mass with $\beta=1\times 10^{-3}$ and $\beta =1$ with a higher resolution. }\label{figure6}
\end{figure*}

We then use the three structure-preserving numerical simulations to investigate the effect of $\beta$ (a parameter related to the lengthscale) on the exchange of mass and energy between the bulk and surface. In the following simulations, we set $M_{23}^{(s)}=0$ and same initial condition as above. To systematically examine the effect of varying $\beta$, we define the definitions of total mass, bulk mass and surface mass as $\beta \int_\Omega \phi d\bx+\int_{\partial \Omega} \phi d\bs$, $\beta \int_\Omega \phi d\bx$ and $\int_{\partial \Omega} \phi d\bs$, without loss the generality. Note that $\beta$ , being related to the characteristic length scale, directly influences the bulk mass contribution. Figure~\ref{figure4}(a–c) displays the snapshots at $t=7,50,100$. For comparison, we also present results with $\beta=1 \times 10^{-3}$ and $\beta=10$ in Figure~\ref{figure4}(d–f) and (g–i), respectively, at the same time instances. We choose the simulation results with $\beta=1$ as the benchmark. In \ref{figure4}-(d-f), $f_m \approxeq -\mu_b$ on $\Gamma$, which is the absorb boundary for $\phi$ so that the mass is concentrated on the surface, which can also be proved in figure \ref{figure6}-(c). When $\beta=1\times 10^{-3}$, the time evolution of total energy, bulk energy and surface energy are slightly different with those with $\beta =1$ as in figure \ref{figure5}. As we analysis in the section 2, small $\beta \rightarrow 0$ indicates the surface mass should be conserved, that is why the time evolution of surface mass with $\beta=1\times 10^{-3}$ is a constant in figure \ref{figure6}-(b). Since the total mass with $\beta =1\times 10^{-3}$ is conserved in figure \ref{figure6}-(a), so the bulk mass with $\beta=1\times 10^{-3}$ in figure \ref{figure6}-(d-e) is also conserved.  By setting $\beta=10$, the exchange of mass and free energy between the bulk and surface is much faster than the other two cases. The surface mass transfers from the surface to bulk and then performs the phase separation and forms a stick pattern near $\Gamma$ in figure \ref{figure4}-(g-i). The exchange process of mass also is adjoint with the exchange of energy, since the bulk energy with $\beta=10$ in figure \ref{figure5}-(b) increases with time firstly and then dissipates with time later, though both the total energy and surface energy are dissipated all time. The surface energy with $\beta=10$ reaches the steady state after $t=7$ in figure \ref{figure5}-(c) due to the surface mass also reaches the steady state in figure \ref{figure6}-(c), which means the exchanges of both mass and energy reach a detailed balance.  
\\
\section{Conclusion}
\noindent \indent In this paper, we have reconstructed four thermodynamically consistent models to describe the exchange of mass and energy between the bulk and the surface, based on fundamental physical laws—namely, mass conservation, volume conservation, and energy dissipation. These models are used to describe the pure irreversible and irreversible-reversible processes, including the transport and reactive transport system. This framework clarifies the physical meaning of the parameter $\beta$ and elucidates the physical mechanism and mathematical structure of the mobility operator.

We show that $\beta$ is intrinsically linked to the characteristic length scale of the system. In the limit $\beta \rightarrow \infty$, the bulk mass/volume remains constant, consistent with the classical thermodynamic assumption of a large system volume, where surface effects become negligible. Conversely, as $\beta \rightarrow 0$, the surface mass/volume remains constant, and bulk dynamics can be disregarded. In both limiting cases, the exchange of mass and energy between the bulk and the surface can be neglected.

For a small system of finite size, the parameter $\beta$ must assume a moderate value to accurately capture the exchange of mass and energy between the bulk and the surface. This requirement aligns with the theory of nanothermodynamics, which states that the total free energy and mass/volume of such a system comprises separate contributions from the bulk and the surface. By using the mass conservation and volume conservation law, we proved that the phase variable in bulk and on surface should not always be same. The Robin boundary condition $\alpha f_m = \beta \mu_s - \mu_b$ can also be reformulated by  $ f_m = \frac{\beta \mu_s - \mu_b}{\alpha}$, where $f_m$ can be viewed as generalized flux, $\frac{1}{\alpha}$ is the mobility operator and $\beta \mu_s - \mu_b$ is the generalized force. This is a special constitutive relation. By systematically varying $\beta$, we are able to investigate the bulk-surface exchange in the numerical results. Our results indicate that a larger $\beta$ enhances the vigor of mass and energy exchange compared to smaller values. When material transfers from the surface into the bulk, it significantly influences the bulk dynamics.

The conservation laws are often overlooked in the development of thermodynamically consistent phase-field models. In this work, by rigorously incorporating these laws, we not only demonstrate the relationship between the parameter $\beta$ and the characteristic length scale but also reveal that conservation laws act as a constraint on the mobility operator. This insight thereby reduces the parameter freedom in model construction, representing a key innovation of our study. Since both Reynolds number and Knudsen number are related to the characteristic length. This work paves the way for formulating hydrodynamics with dynamic boundary conditions, following the same methodology.

\section*{Acknowledgements}
\noindent \indent Xiaobo Jing's work is supported through
National Natural Science Foundation of China (No.12147165), Jiangsu Provincial Scientific Research Center of Applied Mathematics (No.BK20233002), the Start-up Research Fund of Southeast University (No.RF1028623369).
Qi Wang's work is supported by a South Carolina EPSCOR GAIN-CRP award.
\bibliographystyle{plain}
\bibliography{reference}
\end{document}